\documentclass[11pt, a4paper]{article}
\pdfoutput=1
\usepackage[english]{babel}
\usepackage[utf8]{inputenc}

\usepackage{jheppub}
\usepackage{amsmath}
\usepackage{amssymb}
\usepackage{xcolor}
\usepackage{float}
\usepackage{enumerate}
\usepackage{slashed}
 
\usepackage{latexsym}
\usepackage{mathrsfs}
\usepackage{amsthm}
\usepackage{amstext}

\usepackage{graphicx}
\usepackage{xcolor}
\usepackage{slashed}

\makeatletter
\@addtoreset{subfigure}{row}
\makeatother

\usepackage[rightcaption]{sidecap}
\usepackage{caption}
\usepackage{subcaption}

\theoremstyle{plain}
\newtheorem{thm}{Theorem}[section]

\newtheorem{srule}[thm]{Selection rule}

\newcommand {\beq} {\begin{equation}}
\newcommand {\eeq} {\end{equation}}
\newcommand{\bea}{\begin{eqnarray}}
\newcommand{\eea}{\end{eqnarray}}
\newcommand{\bit}{\begin{itemize}}
\newcommand{\eit}{\end{itemize}}
\def\nl{\nonumber \\}

\def\a{\alpha}

\def\p{\partial}

\def\le{\left(}
\def\ri{\right)}

\def\beq{\begin{equation}}
\def\eeq{\end{equation}}

\title{Renormalization properties of  a Galilean Wess-Zumino model} 

\author[a,b]{Roberto Auzzi,} 
\author[c]{Stefano Baiguera,} 
\author[a,d]{Giuseppe Nardelli}
\author[c]{and Silvia Penati}

\affiliation[a]{Dipartimento di Matematica e Fisica, Universit\`a Cattolica
del Sacro Cuore, 
Via Musei 41, 25121 Brescia, Italy}
\affiliation[b]{INFN Sezione di Perugia, Via A. Pascoli, 06123 Perugia, Italy}
\affiliation[c]{Universit\`a degli studi di Milano Bicocca and INFN, Sezione di Milano - Bicocca, Piazza
della Scienza 3, 20161, Milano, Italy}
\affiliation[d]{TIFPA - INFN, c/o Dipartimento di Fisica, Universit\`a di Trento, 38123 Povo (TN), Italy}

\emailAdd{roberto.auzzi@unicatt.it}
\emailAdd{giuseppe.nardelli@unicatt.it}
\emailAdd{s.baiguera@campus.unimib.it}
\emailAdd{silvia.penati@mib.infn.it}

\abstract{
 We consider a  Galilean ${\cal N} =2$ supersymmetric theory with F-term couplings in $2+1$ dimensions,
obtained by null reduction of a relativistic Wess-Zumino model.
We compute quantum corrections and we
check that, as for the relativistic parent theory, the F-term does not
receive quantum corrections. Even more, we find evidence that the causal structure of the non-relativistic dynamics together with particle number conservation constrain the theory 
to be one-loop exact. }

\begin{document}

\maketitle

\section{Introduction}

Emergent symmetries are a recurring theme in
condensed matter physics:  a new symmetry
may arise in the infrared, even if absent from the microscopic
Hamiltonian, due to the presence of an interacting 
infrared fixed point in the renormalization group flow.
For example, Lorentz symmetry can emerge in graphene
 \cite{Semenoff:1984dq,DiVincenzo-Mele,graphene-rev},
whose low-energy excitations can be described
by massless Dirac fermions, which move with a velocity
that is 300 times smaller than the speed of light.

Supersymmetry (SUSY) is a special symmetry which rotates bosonic in fermionic
degrees of freedom and that has been studied for several decades,
mostly from high energy physicist's perspective.
There are several interesting settings where it appears also as an emerging 
symmetry in condensed matter systems.
For example, superconformal invariance in two dimensions arises in the tricritical
Ising model \cite{Friedan:1984rv}. Supersymmetry also
appears in the description of quantum phase transitions
at the boundary of topological superconductors \cite{Grover:2013rc},
in optical lattices \cite{Yu:2010zv},
and in many other settings \cite{Huijse:2014ata,Jian:2014pca,Rahmani:2015qpa,Yu:2019opk,Lee:2006if}.

From a condensed-matter perspective,  there are also many motivations for studying
field theories with non-relativistic symmetries.
This includes both systems with or without the non-relativistic 
boost symmetry (i.e. with  Schr\"odinger  or Lifshitz symmetry).
In this work we will focus on the former, which has a richer symmetry content.
Non-relativistic particles in the limit of infinite
scattering length can be described by a Schr\"odinger  conformal field theory
 \cite{Hagen:1972pd,Jackiw:1990mb,Mehen:1999nd}.
 This has applications in nuclear physics e.g. \cite{Kaplan:1998tg},
 in cold atoms \cite{Nishida:2010tm}  and in the quantum Hall effect \cite{Geracie:2014nka}.
It is then a natural question to investigate
non-relativistic incarnations of SUSY, since this kind of symmetry
might be emergent in the infrared of some real world systems.

The study of Galilean supersymmetry may be also interesting
for holography. Almost all the examples of AdS/CFT correspondence
for which the boundary theory has been precisely identified
correspond to supersymmetric theories. Indeed SUSY gives
a strong analytic control on several quantum physical quantities,
which in some cases can be exactly computed. So, in order to 
find the precise holographic dual of a given gravity background
which geometrically realizes the Schr\"odinger symmetry \cite{Son:2008ye},
it may be useful to focus on an explicitly supersymmetric theoretical setting.

Galilean invariance is usually thought as a low-energy approximation of theories with Poincar\'e
 invariance, and as such it can be found by performing the $c \rightarrow \infty $ limit 
  in the corresponding relativistic setting\footnote{When performing this procedure, divergent expressions in the speed of light appear 
 and we need to introduce some subtraction terms via a chemical potential and
  by appropriately rescaling the fields \cite{Jensen:2014wha}.}. 
On the other hand, it is possible to obtain the Galilean group by Discrete Light Cone Quantization (DLCQ), 
which consists in a dimensional reduction along a null direction of a relativistic theory \cite{Duval:1984cj}. 
SUSY extensions of the Galilean algebra were first introduced in 3+1 dimensions \cite{Puzalowski:1978rv}, where two super-Galilean algebras were constructed, $ \mathcal{S}_1 \mathcal{G} $ which includes a single two-component spinorial supercharge and 
$ \mathcal{S}_2 \mathcal{G} $, which contains two supercharges. They can be obtained as the non-relativistic limit of ${\cal N}=1$ and ${\cal N}=2$ Super-Poincar\'e algebras, respectively. Alternatively, $\mathcal{S}_2 \mathcal{G}$ can be obtained performing a null reduction of the super-Poincar\`e algebra in 4+1 dimensions. It turns out that $ \mathcal{S}_1 \mathcal{G} \subset \mathcal{S}_2 \mathcal{G}$.
 
In 3+1 dimensions theories with  $ \mathcal{S}_1 \mathcal{G} $  and $\mathcal{S}_2 \mathcal{G}$ invariance have been considered in \cite{Puzalowski:1978rv,Clark:1983ne,deAzcarraga:1991fa,Meyer:2017zfg}, 
while in 2+1 dimensions Chern-Simons theories with   $ \mathcal{S}_2 \mathcal{G} $ symmetry
were studied in  \cite{Meyer:2017zfg,Leblanc:1992wu,Bergman:1995zr}. Moreover, SUSY generalizations of the 
Schr\"odinger  algebra have been investigated \cite{Beckers:1986ty,Gauntlett:1990nk,Duval:1993hs,Leblanc:1992wu}, as well as Lifshitz SUSY \cite{Chapman:2015wha}.

In this paper we will build  an example of a theory with $ \mathcal{S}_2 \mathcal{G} $
SUSY in $2+1$ dimensions, which we obtain by null reduction from
a $3+1$ dimensional ${\cal N}=1$ Wess-Zumino model.
 A non-vanishing interaction  in the superpotential which survives the null reduction
requires the introduction of at least two chiral fields. 
Moreover, some exotic derivative couplings emerge when integrating out
the non-dynamical components of the fields. Similar derivative interactions were
recently considered in a $1+1$ dimensional example (without supersymmetry) \cite{Yong:2017ubf}.

One may wonder if  the powerful non-renormalization theorem for the $F$-term
survives null-reduction, giving an interacting
Galilean theory with a nice ultraviolet behaviour. We find that 
the non-relativistic truncation has even better UV properties:
\begin{itemize}
 \item  like the relativistic parent, the model is renormalizable and the superpotential 
 term does not acquire quantum corrections,
 \item there is strong evidence that the whole renormalization 
 of the two-point function is just at one loop (we check this claim
 explicitly up to four loops and discuss in general higher orders).
  This remarkable property is due to the $U(1)$  
 symmetry associated to the non-relativistic particle number conservation,
 which limits the number of diagrams at a given
 perturbative order. Moreover, the causal structure 
strongly reduces the number of non-vanishing diagrams.
 \end{itemize}
As a consequence of these two properties, we work out a set of  selection
rules for diagrams which simplifies the computation of the quantum corrections.

The paper is organised as follows. In section \ref{sect-alg} we derive the 
  $ \mathcal{S}_2 \mathcal{G} $ algebra from null reduction
  and we discuss the ${\cal N}=2$ non-relativistic superspace.
  In section \ref{section-WZ-rel} we briefly  review the relativistic Wess-Zumino model
  and its renormalisation properties.  Sections \ref{sec-Non-relativistic Wess-Zumino model} 
  and \ref{Renormalization in superspace} contain the main results of the paper: 
  we introduce the model and we study its quantum corrections using
  selection rules that we derive in the supergraph formalism.
  We discuss our results and possible developments in section \ref{Conclusions}. Conventions are listed in appendix \ref{app-conv}.
 For completeness, the model in component field formalism and its quantization are discussed in appendices \ref{sec_components} and \ref{app-nonrel WZ model in components}, respectively, while
 in appendix \ref{examples} we present in details an example of supergraph calculation.


\section{Non-relativistic supersymmetry algebra}
\label{sect-alg}

We are interested in studying non-relativistic SUSY theories  in $2+1$ dimensions with $ \mathcal{S}_2 \mathcal{G} $ invariance. This graded generalization of the Galilean algebra contains two complex supercharges and is described by the following non-vanishing (anti)commutators
\begin{eqnarray}
& [P_j, K_k] = i \delta_{jk} M \, , \qquad
[H, K_j] = i P_j \, ,  \nonumber \\ 
& [P_j, J] = -i \epsilon_{jk} P_k \, , \qquad
[K_j, J] = -i \epsilon_{jk} K_k  \, ,  \qquad \qquad j,k=1,2  \label{Bargmann} \\
& \nonumber \\
& [Q,J] = \frac12 Q \, , \qquad
\lbrace Q, Q^{\dagger}  \rbrace = \sqrt{2} M \, , \nonumber \\
& [\tilde{Q},J] = - \frac12 \tilde{Q} \, , \quad
[\tilde{Q}, K_1 - i K_2] = -i Q \, , \quad
\lbrace \tilde{Q}, \tilde{Q}^{\dagger}  \rbrace = \sqrt{2} H \, ,   \nonumber \\
& \lbrace Q, \tilde{Q}^{\dagger} \rbrace = -  (P_1 -i P_2) \, , \quad
\lbrace \tilde{Q}, Q^{\dagger} \rbrace = -  (P_1 + i P_2) 
\label{commu superGalileo2}
\end{eqnarray}
Here $P_j$ are the spatial components of the momentum, $ K_j $ are the generators of Galilean boosts, $ J $ is the planar angular momentum and $Q, \tilde{Q}$ are two complex supercharges. The central charge $M$ corresponds to the mass or particle number conservation.

This is the non-relativistic $ \mathcal{N}=2 $ SUSY algebra in (2+1) dimensions. It first appeared in the non-relativistic SUSY extension of Chern-Simons matter systems, which exhibit an enhanced superconformal symmetry \cite{Leblanc:1992wu}. Its bosonic part is the $U(1)$ central extension of the Galilei algebra, known as Bargmann algebra. Instead, removing $\tilde{Q}$ from \eqref{commu superGalileo2} we obtain the $ \mathcal{S}_1 \mathcal{G}$ algebra.

The  $\mathcal{S}_2 \mathcal{G}$ algebra can be obtained in different ways. We can start with the non-SUSY non-relativistic Galilean algebra, add two supercharges and impose consistency conditions (as done in $3+1$ dimensions \cite{Puzalowski:1978rv}). Alternatively, we can perform the In\"{o}n\"{u}-Wigner contraction of the $2+1$ super-Poincar\`e algebra in the $c \to \infty$ limit \cite{Bergshoeff:2015uaa}. Finally, it can be obtained by null reduction of ${\cal N}=1$ super-Poincar\'e in $3+1$ dimensions. We will follow the last approach, as it is the most convenient one for constructing the non-relativistic ${\cal N}=2$ superspace.


\subsection{Null reduction of relativistic $ \mathcal{N}=1 $ SUSY algebra in 3+1 dimensions}

 We begin by proving that the $ \mathcal{S}_2 \mathcal{G} $ algebra in (\ref{Bargmann},\ref{commu superGalileo2}) can be obtained by null reduction of the relativistic $ \mathcal{N}=1 $ SUSY algebra in $3+1$ dimensions.

Given the (3+1)-dimensional Minkowski spacetime described by light-cone coordinates
\beq\label{light-cone}
x^M = (x^-, x^+ , x^1, x^2)  \equiv (x^- , x^{\mu})  \qquad \quad x^{\pm} = \frac{x^3 \pm x^0}{\sqrt{2}} 
\eeq
null reduction is realized by compactifying $x^-$ on a small circle of radius $ R$.
For convenience,  we rescale $x^- \to x^-/R$ in such a way that the rescaled 
coordinate is adimensional. In order to keep the metric tensor adimensional,
we also rescale $x^+ \to R \, x^+$.

It is well known that the bosonic part of the super-Poincar\'e algebra reduces to the Bargmann algebra \eqref{Bargmann} 
by identifying some components of the linear and angular momenta with the central charge and the boost operator \cite{Duval:1984cj,Son:2008ye}.

To perform the reduction of the fermionic part of the algebra we rewrite the r.h.s. of the four-dimensional anticommutator 
$\{\mathcal{Q}_\alpha , \bar{\mathcal{Q}}_{\dot{\beta}} \} = i \sigma^M_{\alpha \dot{\beta}} \partial_M$
 in terms of light-cone derivatives $\partial_{\pm} = \frac{1}{\sqrt{2}}( \partial_3 \pm \partial_0)$ \footnote{For conventions on four-dimensional spinors see appendix \ref{app-conv}.}
\beq
\lbrace \mathcal{Q} , \bar{\mathcal{Q}} \rbrace = i \begin{pmatrix}
\sqrt{2} \p_+  &   \p_1 - i \p_2 \\
 \p_1 + i \p_2 & - \sqrt{2} \p_{-} 
\end{pmatrix} 
\label{SUSY algebra relativistica}
\eeq 

When the derivatives act on local functions $ \phi (x^M) $ of the (3+1) space-time, 
we set $ \phi (x^M) = e^{i m x^-} \varphi(x^{\mu})$, where \emph{m} is an adimensional parameter. 
Therefore, identifying
\beq
\p_+ \rightarrow \p_t \, , \qquad
\p_- \rightarrow i m 
\eeq 
and reinterpreting the four-dimensional two-spinor components as three-dimensional complex Grassmann scalars with
 $\mathcal{Q}_{\alpha} \to Q_\alpha$, 
 $\bar{\mathcal{Q}}_{\dot{\beta}} \to Q_\beta^\dagger$, we obtain 
\beq
\begin{aligned}
& \lbrace Q_{1} , Q_1^{\dagger} \rbrace =  \sqrt{2} i \p_{t} =   \sqrt{2}  H \, , \qquad \qquad \qquad \quad
 \lbrace Q_{1} , Q_{2}^{\dagger} \rbrace =   i (\p_{1}-i\p_2) = - (P_1 - i P_2)  \, ,  \\
& \lbrace Q_{2} , Q_{1}^{\dagger} \rbrace =   i (\p_{1}+i\p_2) = - (P_1 + i P_2)  \, , \qquad 
\lbrace Q_{2} , Q_{2}^{\dagger} \rbrace = -  \sqrt{2} i \p_{-} =    \sqrt{2} m  
\end{aligned} 
\label{SUSY algebra dalla DLCQ}
\eeq
These anticommutators coincide with the ones in \eqref{commu superGalileo2} if we identify\footnote{We note that identification \eqref{identification} is required to obtain the correct anticommutators $\{ Q,  Q^\dagger \}$ and  $\{ \tilde{Q},\tilde{Q}^\dagger \}$, but it interchanges  $ (P_1 + i P_2) $ and $ (P_1 - i P_2) $ in the mixed anticommutators. This is simply due to the fact that we chose $x^-$ as the compact light-cone coordinate. Had we chosen $x^+$ we would have obtained exactly the algebra in \eqref{commu superGalileo2}. Since having $ (P_1 + i P_2) $ and $ (P_1 - i P_2) $ interchanged does not affect our construction, we take $ x^- $ as the compact direction being this a more conventional choice in the literature.}
\beq
\label{identification}
Q_1 =  \tilde{Q} \, , \qquad
Q_1^{\dagger} =  \tilde{Q}^{\dagger} \, , \qquad
Q_2 = Q \, , \qquad
Q_2^{\dagger} = Q^{\dagger} 
\eeq
and $m$ with the eigeinvalue of the U(1) generator $M$.

It is interesting to compare our result with the relativistic $ \mathcal{N}=2 $ SUSY algebra in (2+1) dimensions that can be obtained via Kaluza-Klein reduction of the $ \mathcal{N}=1 $ SUSY algebra in (3+1) dimensions.
If we start from $\{ \mathcal{Q}_\alpha, \bar{\mathcal{Q}}_{\dot{\beta}} \} = -\sigma^M_{\alpha \dot{\beta}} P_M$, 
compactify along the $ x^3 $ direction and assign momentum $p_3 \equiv Z$, we obtain the three-dimensional anticommutator \cite{Aharony:1997bx}
\beq
\lbrace \mathcal{Q}_{\alpha} , \mathcal{Q}_{\beta}^\dagger \rbrace =  -\sigma^{\mu}_{\,\, \alpha \beta} P_{\mu} +  i \epsilon_{\alpha \beta} Z 
\eeq 
where $ \mu=0,1,2 $ and \emph{Z} plays the role of a central term.
This expression is very similar to the one found in the non-relativistic case, eq. \eqref{SUSY algebra dalla DLCQ} with $m$ playing the role of a central charge. However, while in the relativistic reduction a central term appears in the fermionic part of the algebra when we reduce the number of dimensions, in the non-relativistic case the central charge is produced already in the bosonic sector (without requiring any SUSY extension) and accounts for the physical fact that in non-relativistic theories the particle number is a conserved quantity.

\subsection{Non-relativistic superspace}
\label{section-Non relativistic superfields}

In the relativistic setting the construction of SUSY invariant actions and the study of renormalization properties is better performed in superspace, where fields belonging to the same multiplet are organized in superfields. Having in mind to apply the same techniques to non-relativistic SUSY systems, we first
construct  the ${\cal N}=2$ non-relativistic superspace by applying null reduction to the four-dimensional one. The non-relativistic superspace was first introduced in four dimensions \cite{Clark:1983ne,deAzcarraga:1991fa}, whereas previous constructions in three dimensions based on different techniques can be found in \cite{Bergman:1995zr,Nakayama:2009ku}.
 
We start with the relativistic ${\cal N}=1$ superspace in (3+1) described by superspace coordinates $(x^\mu, \theta^\alpha, \bar{\theta}^{\dot{\alpha}})$. An explicit realization of the super-Poincar\'e algebra is given in terms of the following supercharges\footnote{Superspace conventions are discussed in Appendix \ref{app-conv}.}
\beq
{\cal Q}_{\alpha} = i \frac{\p}{\p \theta^{\alpha}} - \frac12 \bar{\theta}^{\dot{\beta}}  \p_{\alpha \dot{\beta}}  \, , \qquad
\bar{\cal Q}_{\dot{\alpha}} = - i \frac{\p}{\p \bar{\theta}^{\dot{\alpha}}} + \frac12 \theta^{\beta} \p_{\beta \dot{\alpha}} 
\label{supercariche rel 4d}
\eeq
and SUSY covariant derivatives 
\beq
{\cal D}_{\alpha} = \frac{\p}{\p \theta^{\alpha}} - \frac{i}{2} \bar{\theta}^{\dot{\beta}} \p_{\alpha \dot{\beta}}   \, , \qquad
\bar{\cal D}_{\dot{\alpha}} =  \frac{\p}{\p \bar{\theta}^{\dot{\alpha}}} - \frac{i}{2} \theta^{\beta}  \p_{\beta \dot{\alpha}} 
\label{derivate covarianti rel 4d}
\eeq
which act on local superfields $\Psi(x^M, \theta^\alpha, \bar{\theta}^{\dot{\alpha}})$.

The ${\cal N}=2$ non-relativistic superspace in (2+1) dimensions can be easily obtained by suitably generalizing the DLCQ procedure. To this end, we move to light-cone coordinates \eqref{light-cone} and rewrite $\p_{\alpha \dot{\beta}}  = \sigma^M_{\alpha \dot{\beta}} \p_M$ in (\ref{supercariche rel 4d}, \ref{derivate covarianti rel 4d}) in terms of $\p_{\pm}, \p_1,\p_2$. 
Then, mimicking what we have done in the non-supersymmetric case, we reduce a generic four-dimensional field as $ \phi (x^M) = e^{i m x^-} \varphi(x^{\mu})$. Since supersymmetry requires each field component of a multiplet to be an eigenfunction of the $\p_-$ operator with the same eigenvalue $m$, the reduction can be done directly at the level of superfields, by writing
\beq
\Psi(x^M, \theta^\alpha, \bar{\theta}^{\dot{\alpha}}) = e^{im x^{-}} \tilde{\Psi}(x^+ \equiv t, x^{i}, \theta^{\alpha}, (\theta^\alpha)^\dagger) 
\label{decomposition rel and non-rel superfield}
\eeq
Acting on these superfields with supercharges and covariant derivatives (\ref{supercariche rel 4d},\ref{derivate covarianti rel 4d}) rewritten in terms of light-cone derivatives, and performing the identification $\p_+ \equiv \p_t$ and $\p_- \equiv iM$ (with eigenvalue  $m$), we obtain\footnote{From now on we rename $(\theta^\a)^\dagger \equiv \bar{\theta}^\a$ and similarly for the other grassmannian quantities.}
\beq  \label{nonrelQD}
\begin{cases}
Q_1 = i \frac{\p}{\p \theta^1} - \frac12  \bar{\theta}^2 (\p_1 - i \p_2) - \frac{1}{\sqrt{2}} \bar{\theta}^1  \p_t \\
\bar{Q}_1 = -i \frac{\p}{\p \bar{\theta}^1} + \frac12 \theta^2 (\p_1 + i \p_2) + \frac{1}{\sqrt{2}} \theta^1 \p_t \\
Q_2 = i \frac{\p}{\p \theta^2} - \frac12  \bar{\theta}^1 (\p_1 + i \p_2) - \frac{i}{\sqrt{2}} \bar{\theta}^2 M \\
\bar{Q}_2 = -i  \frac{\p}{\p \bar{\theta}^2} + \frac12 \theta^1 (\p_1 - i \p_2) - \frac{i}{\sqrt{2}} \theta^2 M  \\
\end{cases}
\begin{cases}
D_1 =  \frac{\p}{\p \theta^1} - \frac{i}{2} \bar{\theta}^2 (\p_1 - i \p_2) - \frac{i}{\sqrt{2}} \bar{\theta}^1 \p_t \\
\bar{D}_1 =   \frac{\p}{\p \bar{\theta}^{1}} - \frac{i}{2} \theta^2 (\p_1 + i \p_2) - \frac{i}{\sqrt{2}} \theta^1 \p_t \\
D_2 =  \frac{\p}{\p \theta^2} - \frac{i}{2} \bar{\theta}^1 (\p_1 + i \p_2) - \frac{1}{\sqrt{2}} \bar{\theta}^2 M \\
\bar{D}_2 =   \frac{\p}{\p \bar{\theta}^2} - \frac{i}{2} \theta^1 (\p_1 - i \p_2) - \frac{1}{\sqrt{2}} \theta^2 M  \\
\end{cases}
\eeq
These operators realize a representation of the non-relativistic algebra \eqref{commu superGalileo2} and can be interpreted as the supercharges and the covariant derivatives of a three-dimensional ${\cal N}=2$ superspace described by coordinates $(t, x^1, x^2,  \theta^1, \theta^2, \bar{\theta}^1,  \bar{\theta}^2)$. Correspondingly, the functions $\tilde{\Psi}$ in \eqref{decomposition rel and non-rel superfield} are three-dimensional ${\cal N}=2$ superfields realizing a representation of the non-relativistic SUSY algebra. 

We point out that the non-relativistic superspace and the corresponding supercharges could be alternatively constructed by quotienting the SUSY extension of the Bargmann algebra by the subgroup of spatial rotations and Galilean boosts, in analogy with the construction of the relativistic superspace as the quotient super-Poincar\'e$/SO(1,3)$. However, the null reduction procedure is more convenient, as it relies on the quotient algebra already implemented in four dimensions.  

As in the relativistic case, imposing suitable constraints we can reduce the number of superfield components and realize irreducible representations of the superalgebra. In particular, we are interested in the construction of (anti)chiral superfields. These can be obtained either by null reduction of the four-dimensional (anti)chiral superfields,  $\bar{\cal D}_{\dot{\alpha}} \Sigma = 0$ (${\cal D}_{\alpha} \bar{\Sigma} = 0$), or directly in three-dimensional superspace by imposing 
\beq
\bar{D}_\a \Sigma = 0 , \qquad \qquad D_\a \bar{\Sigma} = 0 
\eeq
where the covariant derivatives are given in \eqref{nonrelQD}. 

Defining coordinates 
\beq
x_{L,R}^{\mu} = x^{\mu} \mp i \theta^{\alpha} (\bar{\sigma}^\mu)_{\a \beta} \bar{\theta}^{\beta}  \qquad \qquad \mu = +, 1,2
\eeq
which satisfy $\bar{D}_\a x_L^\mu =0, D_\a x_R^\mu =0$, the (anti)chiral superfields have the following expansion
\begin{align} 
\label{supercampi x_L x_R}
& \Sigma(x_L^\mu, \theta^{\alpha}) = \varphi(x_L^\mu) +  \theta^{\alpha} \tilde \psi_{\alpha} (x_L^\mu) - \theta^2 F(x_L^\mu) & \\
& \bar{\Sigma}(x_R^\mu,  \bar{\theta}^{\beta}) = \bar{\varphi}(x_R^\mu) +  \bar{\theta}_{\gamma} \Bar{\tilde \psi}^{\gamma} (x_R^\mu) - \bar{\theta}^2 \bar{F}(x_R^\mu) &
\end{align}
 
Manifestly supersymmetric actions can be constructed by using the Berezin integral on spinorial coordinates. In the relativistic superspace, for a generic superfield $\Psi$ we define
\beq \label{relberezin}
\int d^4x d^4\theta \, \Psi = \int d^4x \, {\cal D}^2 \bar{\cal D}^2 \Psi \Big|_{\theta = \bar{\theta}=0}
\eeq
with covariant derivatives given in \eqref{derivate covarianti rel 4d}. 
Performing the null reduction and extracting the $x^-$ dependence of the superfield by setting $\Psi = e^{imx^-} \tilde{\Psi}$, we obtain the prescription for the Berezin integrals in the non-relativistic superspace
\beq \label{Berezin integration null reduction}
\begin{aligned}
& \int d^4x d^4\theta \, \Psi = \int d^4x \, {\cal D}^2 \bar{\cal D}^2 \Psi \Big|_{\theta = \bar{\theta}=0} \;  \longrightarrow  \\
&  \int d^3x D^2 \bar{D}^2 \tilde{\Psi} \Big|_{\theta = \bar{\theta}=0} \; \times \frac{1}{2\pi} \int_0^{2\pi}  dx^- \,  e^{imx^-}   \equiv \int d^3x d^4\theta \, \tilde{\Psi}  \; \times \frac{1}{2\pi} \int_0^{2\pi}  dx^- \,  e^{imx^-}
\end{aligned}
\eeq
where in the r.h.s. $d^3x \equiv dt \,dx^1 dx^2$ and the spinorial derivatives are the ones in eq. \eqref{nonrelQD}. 
It is immediate to observe that whenever $m \neq 0$ we obtain a trivial reduction due to the $x^-$ integral. Non-vanishing expressions arise only if the super-integrand $\Psi$ is uncharged respect to the mass generator. In the construction of SUSY invariant actions this is equivalent to require the action to be invariant under one extra global U(1) symmetry \cite{deAzcarraga:1991fa}.

 
\section{Review of the relativistic Wess-Zumino model}
\label{section-WZ-rel}

In this Section we briefly review the renormalization of the relativistic four-dimensional Wess-Zumino (WZ) model, both in superspace and in components, in order to fix our notations and recall the main physical properties that we plan to investigate in a non-relativistic set-up.

The classical action of the WZ model \cite{Wess:1974tw} in (3+1) dimensions is given by 
\beq
S= \int d^4 x \, d^4 \theta  \,  \bar{\Sigma} \Sigma + \int d^4 x \, d^2 \theta \, \le \frac{m}{2} \Sigma^2  + \frac{\lambda}{3!} \Sigma^3  \ri  + \mathrm{h.c.}
\label{classical rel WZ model}
\eeq
and describes the dynamics of the field components of a chiral superfield $\Sigma = (\phi, \psi, F)$. For simplicity we focus on the massless model, so from now on we set $m=0$.  This model is classically scale invariant.

When reduced in components using definitions \eqref{components} the action reads 
\beq
S =  
\int d^4x \Big[ - \p^M \bar{\phi} \, \p_M \phi + i \bar{\psi}  \bar{\sigma}^M \p_M \psi + \bar{F}  F   
+ \left( 3 \lambda F \phi^2 - 3 \lambda \psi^{\alpha} \psi_{\alpha} \phi + \mathrm{h.c.} \right) \Big]
\label{component action}
\eeq
The action in \eqref{classical rel WZ model} is manifestly invariant under ${\cal N}=1$ SUSY transformations
\beq
\delta_{\varepsilon} \Sigma = \left[i \varepsilon^{\alpha} {\cal Q}_{\alpha} + i \bar{\varepsilon}_{\dot{\alpha}} \bar{\cal Q}^{\dot{\alpha}}  , \Sigma \right] 
\eeq
Equivalently, action \eqref{component action} is invariant under
\beq
\begin{cases}
\delta_{\varepsilon} \phi =  -\varepsilon^{\alpha} \psi_{\alpha} \\
\delta_{\varepsilon} \psi_{\alpha} =  i \bar{\varepsilon}^{\dot{\alpha}} (\p_{\alpha {\dot{\alpha}}} \phi) + \varepsilon_{\alpha} F \\
\delta_{\varepsilon} F =  - i \bar{\varepsilon}^{\dot{\alpha}} \p_{\alpha {\dot{\alpha}}} \psi^{\alpha} 
\label{Rel SUSY variations in components}
\end{cases}
\eeq

\subsection{Renormalization in superspace}
\label{sect-Quantum Wess-Zumino model and supergraph formalism}

At quantum level we consider the generating functional  
\beq
Z [J, \bar{J}] = \int \mathcal{D} \Sigma \, \mathcal{D}  \bar{\Sigma}  \, \exp \left\lbrace i \le S + \int d^2 \theta \, J \Sigma + \int d^2 \bar{\theta} \, \bar{J}  \bar{\Sigma} \ri  \right\rbrace 
\eeq
where the sources $ J, \bar{J} $ are chiral and anti-chiral superfields, respectively. Correlation functions can be obtained by repeated application of functional derivatives 
\beq\label{sources}
\frac{\delta J (z_i)}{\delta J(z_j)} =  \bar{\cal D}^2 \, \delta^{(8)} (z_i - z_j) \, , \qquad
\frac{\delta \bar{J} (z_i)}{\delta \bar{J}(z_j)} =  {\cal D}^2  \, \delta^{(8)} (z_i - z_j) 
\eeq
where $ z \equiv (x^M, \theta^{\alpha}, \bar{\theta}^{\dot{\alpha}}) $ and $\delta^{(8)} (z_i - z_j) \equiv \delta^{(4)} (x_i - x_j) \delta^{(2)} (\theta_i - \theta_j)\delta^{(2)} (\bar{\theta}_i - \bar{\theta}_j)$. The additional covariant derivatives acting on the delta functions come from the fact that we are deriving constrained superfields.

Renormalizability properties can be investigated  in superspace, where Feynman rules can be formulated directly for superfields. These allow to draw supergraphs which can be eventually reduced to ordinary Feynman integrals by performing D-algebra.

In short, for the massless WZ model super-Feynman rules are \cite{Gates:1983nr} 
\begin{itemize}
\item
Superfield propagator  
\beq
\langle \Sigma (z_i) \bar{\Sigma} (z_j) \rangle =  \frac{1}{\square} \, \delta^{(8)} (z_i - z_j) \; \longrightarrow  \; 
\langle \Sigma (p) \bar{\Sigma} (-p) \rangle = - \frac{1}{p^2} \, \delta^{(4)} (\theta_i - \theta_j)
 \label{relativistic superpropagators WZ model}
\eeq 

\item
Vertices. These are read directly from the interaction Lagrangian. They are cubic vertices containing only chiral or anti-chiral superfields.
Because of identity \eqref{sources} we assign one $  \bar{\cal D}^2 $ ($ {\cal D}^2 $) to every internal line exiting from a chiral (anti-chiral) vertex. One of these factors is then used to complete the chiral (anti-chiral) integral at the vertex, thus dealing only with $ \int d^4 \theta $ at each vertex. 

\end{itemize}

At this point we need to perform the spinorial integrals exploiting the spinorial delta functions, in order to obtain a final result which is a local function of $(\theta, \bar{\theta})$ integrated in $d^4\theta$. However, we need to take into account that spinorial deltas may be partially affected by residual ${\cal D}$'s or $\bar{\cal D}$'s acting on internal lines. Moreover, products of identical deltas are subject to (we set $ \delta_{ij} \equiv \delta^{(2)} (\theta_i - \theta_j) \, \delta^{(2)} (\bar{\theta}_i - \bar{\theta}_j ) $) 
\begin{align}
& \delta_{ij} \delta_{ij} = 0 \, , \qquad
\delta_{ij} {\cal D}^{\alpha} \delta_{ij} = 0 \, , \qquad
\delta_{ij} {\cal D}^{2} \delta_{ij} = 0 \, , \qquad
\delta_{ij} {\cal D}^{\alpha} \bar{\cal D}^{\dot{\alpha}} \delta_{ij} = 0 \, , \qquad \delta_{ij} {\cal D}^{\alpha} \bar{\cal D}^2 \delta_{ij} = 0
 &  \nonumber \\
 &  \nonumber \\
 &   \delta_{ij} {\cal D}^{\alpha} \bar{\cal D}^2 {\cal D}^{\beta} \delta_{ij} = -\epsilon^{\alpha \beta} \delta_{ij} \, , \qquad   \delta_{ij} {\cal D}^{2} \bar{\cal D}^2  \delta_{ij} = \delta_{ij} \bar{\cal D}^2 {\cal D}^{2} \delta_{ij} =  \delta_{ij} \frac{ {\cal D}^{\alpha} \bar{\cal D}^2 \mathcal{D}_{\alpha}}{2}  \delta_{ij} =  \delta_{ij}   &
 \label{rules covariant derivatives on delta functions}
\end{align}
Therefore, it is easy to verify that we need to perform D-algebra until we reach a configuration in which exactly two ${\cal D}$'s and two $\bar{\cal D}$'s survive in each loop. 
This amounts to integrate by parts spinorial derivatives at the vertices and trade products of them with space-time derivatives through commutation rules like
\beq
[{\cal D}^{\alpha}, \bar{\cal D}^2] = i \p^{\alpha \dot{\alpha}} \bar{\cal D}_{\dot{\alpha}} \, , \quad [\bar{\cal D}^{\dot{\alpha}}, {\cal D}^2] = -i \p^{\alpha \dot{\alpha}} {\cal D}_{\alpha} \, , \quad
{\cal D}^2 \bar{\cal D}^2 {\cal D}^2 = \square {\cal D}^2 \, , \quad  \bar{\cal D}^2 {\cal D}^2 \bar{\cal D} ^2 = \square \bar{\cal D}^2
\label{rules covariant derivatives giving momenta in supergraphs}
\eeq
Whenever in a loop we end up with a number of derivatives which less than $2{\cal D}$'s $+ 2\bar{\cal D}$'s the configuration vanishes and can be discharged. Instead, when in a loop we are left with two ${\cal D}$'s plus two $\bar{\cal D}$'s the spinorial integrations associated to that loop can be performed and we are left with a non-vanishing expression local in the spinorial coordinates. 

In so doing, we reduce a supergraph to the sum of a number of ordinary Feynman diagrams. As usual, in momentum space these correspond to integrals over loop momenta, with momentum conserved at each vertex. In general UV and IR divergences arise, which require suitable regularizations to perform the integrals. At the end of the calculation, going back to configuration space we obtain contributions that are given by local functions of the superspace coordinates integrated in $d^4x d^4\theta$. 

\vskip 10pt
The WZ model is renormalizable by power counting. Applying the supergraph techniques described above, it immediately follows that UV divergences always arise in the form of non-chiral superspace integrals and, as such, can only contribute to the kinetic part of the effective action. The cubic superpotential, at the contrary, never gets divergent corrections,  and consequently it does not undergo any renormalization. This is the proof of the well-known perturbative non-renormalization theorem  \cite{Grisaru:1979wc}. 
 
Cancellation of loop divergences requires a wavefunction renormalization. Due to the non-renormalization theorem, the coupling constant of the model inherits a non-trivial renormalization as well. In fact,  
\beq
\mathcal{L} = \int d^4 \theta \, (\bar{\Sigma} \Sigma) + \int d^2 \theta \, (\lambda \Sigma^3) \, \rightarrow
\mathcal{L}_{\rm ren} = \int d^4 \theta \, Z_{\Sigma} (\bar{\Sigma} \Sigma) + \int d^2 \theta \, Z_{\lambda} Z_{\Sigma}^{3/2} (\lambda \Sigma^3)  
\eeq
but the absence of chiral divergences implies 
\beq
Z_{\lambda} Z_{\Sigma}^{3/2} = 1 \; \; \Longrightarrow \; \;
Z_{\lambda} = Z_{\Sigma}^{-3/2} 
\label{statement non-renormalization theorem on Z}
\eeq

\subsection{Renormalization in components}
\label{sect-Renormalization using the component fields formulation}

It is interesting to see how the non-renormalization theorem is formulated when we perform perturbative calculations in components. In view of the comparison with the non-relativistic analysis, this is motivated by the fact that so far most of the literature on non-relativistic systems has used the component formalism. 
 
Starting with the action in components given in \eqref{component action} we can easily obtain the corresponding Feynman rules. If we do not eliminate the auxiliary fields the propagators are the ordinary scalar and fermion propagators completed with $\langle F \bar{F} \rangle = 1$ \footnote{They can also be obtained by reducing the super-propagator \eqref{relativistic superpropagators WZ model} in components.}, while the vertices are still cubic vertices, as inferred directly from the action. 

Evaluating ordinary Feynman diagrams and isolating the UV divergent terms, the renormalizability of the model allows to write
\beq
\mathcal{L}_{\mathrm{ren}} =  - Z \, \p^M \bar{\phi} \, \p_M \phi + i Z \bar{\psi} \bar{\sigma}^M \p_M \psi + Z \bar{F} F 
 + \left( 3 \lambda  Z_{\lambda} Z^{3/2} F \phi^2 + 3 \lambda   Z_{\lambda} Z^{3/2} \psi^{\alpha} \psi_{\alpha} \phi + \mathrm{h.c.}
\right)
\eeq
where we have used the SUSY condition $Z_{\phi} = Z_{\psi} = Z_F \equiv Z$. 
It follows that the non-renormalization theorem still leads to condition (\ref{statement non-renormalization theorem on Z}). In fact, since we have not eliminated the auxiliary field $F$, this is nothing but a trivial rephrasing of the superspace approach. 

Instead, we can proceed by first integrating out the auxiliary field \emph{F} from the action, using its equations of motion. Performing the perturbative analysis and taking into account the non-renormalization condition \eqref{statement non-renormalization theorem on Z}, it turns out that the renormalized action for the dynamical fields reads
\beq
\mathcal{L}_{\mathrm{ren}} = - Z \p^M \bar{\phi} \, \p_M \phi + i Z \bar{\psi} \bar{\sigma}^M \p_M \psi + \left( 3 \lambda  \psi^{\alpha} \psi_{\alpha} \phi - 9 Z^{-1} |\lambda|^2 |\phi|^4 + \mathrm{h.c.} \right)
\eeq
The relevant fact is that while the cubic vertex is still non-renormalized, the quartic scalar interaction renormalizes non-trivially, due to the wavefunction renormalization.
This shows that when working in components and integrating out the auxiliary fields, quantum corrections to the vertices may arise, although the non-renormalization theorem is still at work.

\subsection{The non-renormalization theorem}
The holomorphicity of the superpotential is a powerful constraint which forces all quantum corrections
to $F$-terms to vanish.  At perturbative level, a direct proof can be obtained by supergraphs technique, as reviewed above.
The non-perturbative derivation of this result follows instead from an argument due to Seiberg \cite{Seiberg:1993vc}.

Here we quickly review the argument, following \cite{Weinberg:2000cr}.
We consider a WZ model for $n$ chiral superfields $\Sigma_a$ interacting through a generic 
superpotential $W$
\beq
S= \int d^4 x \, d^4 \theta  \,  \bar{\Sigma}_a \Sigma_a + \int d^4 x \, d^2 \theta \,  
W(\Sigma_a) + \mathrm{h.c.} 
\label{WZ-generico}
\eeq
We introduce one extra chiral superfield $Y$, whose scalar part is set to $1$ to recover the original action, whereas the spinorial and auxiliary components vanish identically. We assign R-charges $R(\Sigma_a)=0$ and $R(Y)=2$.
We also introduce real superfields $Z_{ab}$ for the wave function renormalization
\beq
\tilde{S}= \int d^4 x \, d^4 \theta  \, Z_{ab} \bar{\Sigma}_a \Sigma_b + \int d^4 x \, d^2 \theta \,  
Y \, W(\Sigma_a) + \mathrm{h.c.}
\label{WZ-generico2}
\eeq

Assuming that the regularization procedure does not spoil SUSY,  the Wilsonian effective action at a given scale $\lambda$
is of the following form
\beq
\tilde{S}_\lambda= \int d^4 x \, d^4 \theta   \,  K(\bar{\Sigma}_a \Sigma_a,Z_{ab}, Y, \bar{Y}, \mathcal{D})
+\int d^4 x \, d^2 \theta \, W_\lambda (\Sigma_a, Y)  + \mathrm{h.c.}
\label{WZ-generico3}
\eeq
Then R-invariance and holomorphicity of the superpotential force $W_\lambda$ to be of the form
\beq
W_\lambda (\Sigma_a, Y) =Y \,  W_\lambda(\Sigma_a) 
\label{WZ-generico4}
\eeq
Taking the weak coupling limit $Y \rightarrow 0$, the only contribution to the superpotential is a tree-level vertex, and therefore we find $ W_\lambda(\Sigma_a) = W(\Sigma_a)$.


\section{The non-relativistic Wess-Zumino model}
\label{sec-Non-relativistic Wess-Zumino model}

We now study the non-relativistic counterpart of the WZ model using the superfield formulation of section \ref{section-Non relativistic superfields}. We are primarily interested in investigating if and how the renormalization properties of the relativistic model survive in this case. 

The natural way to obtain the non-relativistic version of the WZ model is by applying null reduction \eqref{Berezin integration null reduction} to the action in (\ref{classical rel WZ model}). Setting
$\Sigma = e^{imx^-} \Phi$ there, we immediately see that while the canonical Kahler potential survives the reduction being U(1) neutral, the holomorphic superpotential has charge 3 and is killed by the $x^-$ integration. The only way-out to obtain an interacting non-relativistic scalar model is then to introduce at least two species of superfields with different $m$ charges, and trigger them in such a way that also the superpotential turns out to be neutral.

We then start in four dimensions with a WZ model for two massless fields described by the action 
\beq
 S = \int d^4 xd^4 \theta    \, \le \bar{\Sigma}_1 \Sigma_1 + \bar{\Sigma}_2 \Sigma_2 \ri  + g  \int  d^4 x d^2 \theta    \, \Sigma_1^2 \Sigma_2 + \mathrm{h.c.}
 \label{WZM-before-null-red}
\eeq
We perform the null reduction by setting
\beq
\Sigma_1 (x^M, \theta, \bar{\theta}) = \Phi_1 (x^{\mu}, \theta, \bar{\theta}) \, e^{imx^{-}} \, , \qquad \Sigma_2 (x^M, \theta, \bar{\theta}) = \Phi_2 (x^{\mu}, \theta, \bar{\theta}) \,  e^{-2imx^{-}} \, 
\label{superfields sector 1 and 2 non rel WZ model}
\eeq
so that the superpotential is neutral under the mass generator. The reduced action reads
\beq
\label{non-rel WZ action in superfield formalism} 
S = \int   d^3 x d^4 \theta \le \bar{\Phi}_1 \Phi_1 +  \bar{\Phi}_2 \Phi_2 \ri + g \int   d^3 x  d^2 \theta \,  \Phi_1^2 \Phi_2 + \mathrm{h.c.}
\eeq
We will refer to the superfields in eq. (\ref{superfields sector 1 and 2 non rel WZ model}) as belonging to sector 1 and 2, respectively.  Since in the non-relativistic superspace the time coordinate has twice the dimensions of the spatial ones, superfields have still mass dimension one and the coupling $g$ is dimensionless. Therefore the model shares classical scale invariance with its relativistic counterpart.

This action is invariant under the non-relativistic ${\cal N}=2$ supersymmetry. Using definition \eqref{Berezin integration null reduction} for the non-relativistic spinorial integrals, we can reduce it to components. 

Focusing first on the kinetic part of the action, we can integrate out the auxiliary fields and obtain (for details see appendix \ref{sec_components})
\beq\label{actioncomponents}
\begin{aligned}
S_{kin}  =  \int d^3 x & \, \left[   2 i m \bar{\varphi}_1 \p_t \varphi_1 + \bar{\varphi}_1 \p_i^2 \varphi_1 
- 4 i m \bar{\varphi}_2 \p_t \varphi_2   + \bar{\varphi}_2 \p_i^2 \varphi_2   \right. \\
& \; \left. + 2 i m  \bar{\chi}_1 \p_t \chi_1 +  \bar{\chi}_1 \p_i^2 \chi_1 + 4 i m \bar{\chi}_2 \p_t \chi_2
 -  \bar{\chi}_2 \p_i^2 \chi_2 \right]
\end{aligned}
\eeq  
where $ \varphi_{1,2} $ and $ \chi_{1,2} $ are the dynamical non-relativistic scalar and fermion fields, respectively. 

If we apply Fourier transform 
\beq
 \varphi (x^{\mu}) = \int \frac{ d\omega \, d^2 k }{(2 \pi)^3} \, a (\vec{k}) e^{-i ( \omega t - \vec{k} \cdot \vec{x}) }
\eeq
to both scalars and fermions, the free equations of motion lead to the following dispersion relations 
\beq
\omega_1 =   \frac{\vec{k_1}^2}{2 m} \qquad \qquad \omega_2 = -  \frac{\vec{k_2}^2}{4 m} 
\eeq
The wrong sign for the energy of $\varphi_2$ and $\chi_2$ is due to U(1) invariance which forces to assign a negative eigenvalue to the mass operator for $\Phi_2$ in decomposition (\ref{superfields sector 1 and 2 non rel WZ model}).  

To circumvent this problem we first integrate by parts $\p_t, \p_i^2$ in sector 2, obtaining 
\beq
\begin{aligned}
S_{kin}   = \int d^3 x & \, \left[   2 i m \bar{\varphi}_1 \p_t \varphi_1 + \bar{\varphi}_1 \p_i^2 \varphi_1 
+ 4 i m \varphi_2\p_t \bar{\varphi_2}    + \varphi_2\p_i^2 \bar{\varphi}_2     \right. \\
& \; \left. + 2 i m  \bar{\chi}_1 \p_t \chi_1 +  \bar{\chi}_1 \p_i^2 \chi_1 + 4 i m  \chi_2 \p_t\bar{\chi}_2   
 + \chi_2 \p_i^2 \bar{\chi}_2     \right]
\end{aligned}
\eeq 
Then we interchange the roles of $\varphi_2$ and $\bar{\varphi}_2$ and similarly of $\chi_2$ and $\bar{\chi}_2$. This operation is equivalent to reversing the role of creation and annihilation operators. At the level of superfields this amounts to interchanging all the components of $ \Phi_2$ with the components of $ \bar{\Phi}_2 $.
Note that this operation is done without exchanging the grassmannian coordinates, i.e. without changing the chirality of the superfield. 
From now on we name $ \Phi_2$ the chiral superfield whose components\footnote{ For details about the null reduction of non-relativistic fermions, see Appendix \ref{app-conv}.} are $(\bar{\varphi}_2, \bar{\xi}_2, \bar{\chi}_2, \bar{F}_2)$, while the antichiral $\bar{\Phi}_2$ has components $(\varphi_2, \xi_2, \chi_2, F_2)$, and assign positive mass $2m$ to $\Phi_2$.

Under this exchange and having eliminated the auxiliary fields, the complete action in components reads
\beq
\begin{aligned}
S   =   & \int d^3x  \; \Big[ 2 i m \bar{\varphi}_1 \p_t \varphi_1 + \bar{\varphi}_1 \p_i^2 \varphi_1 
+ 4 i m \bar{\varphi}_2 \p_t \varphi_2  + \bar{\varphi}_2 \p_i^2 \varphi_2  \\
& \quad  + 2 i m  \bar{\chi}_1 \p_t \chi_1 +  \bar{\chi}_1 \p_i^2 \chi_1  + 4 i m \bar{\chi}_2 \p_t \chi_2
 +  \bar{\chi}_2 \p_i^2 \chi_2 - 4 |g|^2 |\varphi_1\varphi_2|^2 -  |g|^2 |\varphi_1|^4   \\
&  -  i g \left( \sqrt{2} \varphi_1 \chi_1 (\p_1 - i \p_2) \bar{\chi}_2 -2 \bar{\varphi}_2 \chi_1 (\p_1 - i \p_2) \chi_1 + 2 \sqrt{2} \varphi_1 ((\p_1 - i \p_2)\chi_1) \bar{\chi}_2  \right) + {\rm h.c.} \\
 & + 2 |g|^2 \left( -|\varphi_1|^2 \bar{\chi}_1 \chi_1 - 4 |\varphi_1|^2 \bar{\chi}_2 \chi_2 + 2 |\varphi_2|^2 \bar{\chi}_1 \chi_1 + 2 \sqrt{2} \varphi_1 \varphi_2 \bar{\chi}_1 \bar{\chi}_2 + 2 \sqrt{2} \bar{\varphi}_1 \bar{\varphi}_2 \chi_2 \chi_1 \right)  \Big]
\end{aligned}   
\label{Lagrangiana finale WZ interagente potenziale cubico}
\eeq
We note the presence of  cubic derivative interactions, together with the standard quartic couplings.
 Similar derivative interactions in a non-supersymmetric $1+1$ dimensional 
 Galilean model have been recently studied in \cite{Yong:2017ubf}.

We remark that the same action could be obtained by null reduction of the $3+1$ relativistic WZ action in components where the auxiliary fields have been integrated out. 
Similar computations were performed in \cite{Auzzi:2017jry}.

\section{Renormalization in superspace}
\label{Renormalization in superspace}

  We now study the renormalization properties of the model defined by eq.
 (\ref{non-rel WZ action in superfield formalism}), using superspace formalism\footnote{
The corresponding computation in component formalism is reported in Appendix \ref{app-nonrel WZ model in components}.}. 
 To this purpose, we first collect all the super-Feynman and selection rules
 which  select the allowed non-vanishing diagrams. 
 This can be done by performing the null reduction of the relativistic rules in 3+1 
dimensions (see Section \ref{sect-Quantum Wess-Zumino model and supergraph formalism}). 
In addition, we have to take into account the $U(1)$ symmetry of the Galilean action 
\eqref{non-rel WZ action in superfield formalism} associated to the mass central charge $M$. 
This implies that the particle number has to be conserved at each vertex and the only non-vanishing
Green functions are the ones whose external particle numbers add up to zero.

\subsection{Super-Feynman and selection rules}

\begin{figure}[h]
\begin{subfigure}[b]{0.5\linewidth}
\centering
\includegraphics[scale=1]{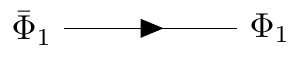}
\end{subfigure}
\begin{subfigure}[b]{0.5\linewidth}
\centering
\includegraphics[scale=1]{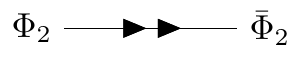}
\end{subfigure}
\begin{subfigure}[b]{0.5\linewidth}
\centering
\includegraphics[scale=0.9]{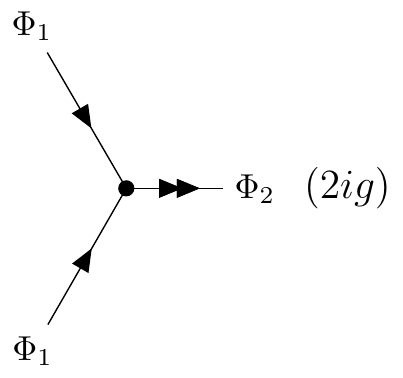}
\end{subfigure}
\begin{subfigure}[b]{0.5\linewidth}
\centering
\includegraphics[scale=0.9]{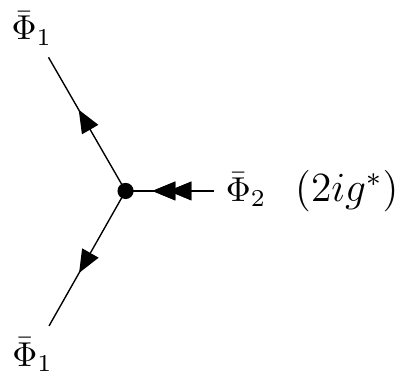}
\end{subfigure}
\caption{Propagators and vertices in superspace.}
\label{fig2-Superpropagators e vertices WZ model nonrel}
\end{figure}

In order to take into account at graphical level the $U(1)$ charge of the two superfields, 
it is useful to indicate the particle number (or mass) flow
by an arrow on the propagator line. 
As shown in fig. \ref{fig2-Superpropagators e vertices WZ model nonrel}
 we assign a single arrow to $\Phi_1$ which has mass $m$ and a double arrow to $\Phi_2$ which has mass $2m$.

The Feynman rules in the non-relativisitic ${\cal N}=2$ superspace are:
\begin{itemize}
\item
Superfield propagators. These are easily obtained from the relativistic ones  in eq. (\ref{relativistic superpropagators WZ model}) by replacing
$\square \rightarrow 2 i M \, \p_t + \p_i^2 $, with $M = m$ or $ 2 m$ or, in momentum space, $ -p^2 \to 2 M \omega - \vec{p}^{\, 2}$. We obtain
\beq
\langle \Phi_1 (\omega, \vec{p}) \bar{\Phi}_1 (-\omega, -\vec{p}) \rangle =  i\frac{ \,  \delta^{(4)} (\theta_1 - \theta_2)  }{2m \omega -\vec{p}^{\, 2} + i \varepsilon}  \, , \quad
\langle \bar{\Phi}_2 (\omega, \vec{p}) \Phi_2 (-\omega, -\vec{p}) \rangle =  i \frac{ \delta^{(4)} (\theta_1 - \theta_2)  }{4m \omega -\vec{p}^{\, 2} + i \varepsilon}  
\label{superpropagatori momentum space}
\eeq
As usual in the Galilean setting, we take the energy dimensions as
\beq
[\omega]= E^2 \, , \qquad [\vec{k}]= E \, , \qquad [m]=E^0 
\eeq

The propagators for both sectors have a retarded $ i \varepsilon $ prescription
which follows the order of fields shown in fig. \ref{fig2-Superpropagators e vertices WZ model nonrel}, where the exchange of particles with anti-particles in sector 2 is manifest from  the reversed order of the fields with respect to sector 1 \footnote{In configuration space the $ i \varepsilon $ prescription translates into a retarded prescription for the propagator. In fact, the Fourier transform of \eqref{superpropagatori momentum space} reads ($M=m$ or $2m$)
\beq
D(\vec{x} , t)= \int \frac{d^2 p \,  d \omega}{(2 \pi)^3} 
i\frac{ \delta^{(4)} (\theta_1 - \theta_2)}{2 M \, \omega - \vec{p}^2 + i \varepsilon} \, e^{-i (\omega t - \vec{p} \cdot \vec{x})} =
 - \frac{i \, \Theta (t)}{4 \pi \,  t} e^{i\frac{M \vec{x}^2}{2t}} \,  \delta^{(4)} (\theta_1 - \theta_2)
\eeq
where $\Theta$ is the Heaviside function.}.

\item 
Vertices. These are cubic vertices easily read from the action in \eqref{non-rel WZ action in superfield formalism}. The particle number conservation at each vertex translates into the condition that the numbers of entering and exiting arrows have to match (see fig. \ref{fig2-Superpropagators e vertices WZ model nonrel}).
\end{itemize}

\vskip 10pt

Since the null reduction does not affect the grassmannian part of the superspace, supergraphs are built as in the relativistic case. In particular, rules \eqref{sources} still hold, so that we have one extra $\bar{D}^2$ ($D^2$) for each chiral (anti-chiral) superfield entering or exiting a vertex. The only important difference is that in the present case the grassmannian derivatives are the non-relativistic ones in \eqref{nonrelQD}. D-algebra can then be performed as summarized in section \ref{sect-Quantum Wess-Zumino model and supergraph formalism} in order to reduce the supergraph to a combination of ordinary Feynman graphs for functions that are local in $(\theta, \bar{\theta})$. In the non-relativistic case identities that are crucial for the D-algebra are still \eqref{rules covariant derivatives on delta functions} (with $D, \bar{D}$ replacing ${\cal D}, \bar{\cal D}$), and  (see eqs. (\ref{3d_derivatives}, \ref{useful_identities}))
\begin{align}
&[D^{\alpha}, \bar{D}^2] = \sqrt{2} M
\bar{D}_1 \delta^{\alpha}_{ 1}  + i (\bar{\sigma}^{\mu})^{\alpha \beta} \p_{\mu} \bar{D}_{\beta} 
 \, , \quad  [\bar{D}^{\alpha}, {D}^2] = -\sqrt{2}  M
D_1 \delta^{\a}_{ 1} - i (\bar{\sigma}^{\mu})^{\alpha \beta} \p_{\mu} D_{\beta}  \, ,  \nonumber & \\ 
& {D}^2 \bar{D}^2 {D}^2 = (2 i M \, \p_t + \p_i^2) D^2  \, , \qquad \qquad  \; \;  \bar{D}^2 {D}^2 \bar{ D} ^2 = (2 i M \, \p_t + \p_i^2) \bar{D}^2 
\label{rules covariant derivatives giving momenta in supergraphs 2}
\end{align}
where $  \mu \in \lbrace +,1,2 \rbrace$.

Since the interaction part of the action still contains cubic vertices as in the relativistic case, the possible topologies of supergraphs are the same (for supergraphs of the ordinary WZ model, see for instance \cite{Abbott:1980jk}, \cite{Sen:1981hk}).
However, the particle number conservation combined with the analyticity properties
leads to extra selection rules that are peculiar of the non-relativistic models, and drastically reduce the number of non-vanishing diagrams. 

\begin{figure}[h]
\centering
\includegraphics[scale=1.4]{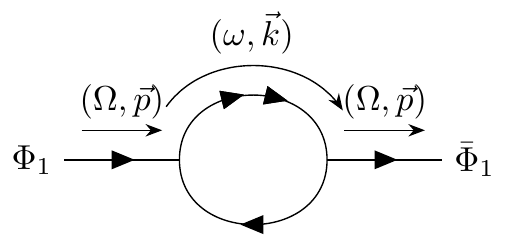}
\caption{One-loop correction to the self-energy of the $\Phi_1$ superfield.}
\label{fig3_rinormalizzione self-energy supercampo 1}
\end{figure}

First of all, the retarded nature of the non-relativistic scalar propagator, which in momentum space is linear in the energy $ \omega$, implies
\begin{srule} 
\label{srule2}
Arrows inside a Feynman diagram cannot form a closed loop.
\end{srule}
This can be easily seen to be a consequence of the residue theorem in momentum space and is better illustrated with an example.
We consider the quantum correction to the self-energy of the superpropagator in sector 1, as depicted in fig. \ref{fig3_rinormalizzione self-energy supercampo 1}.

This diagram gives the following contribution to the effective action 
\beq
i \Gamma_1^{(2)}  (\Phi_1, \bar{\Phi}_1) =  4 |g|^2  \int d^4 \theta \, \frac{d\omega \, d^2k}{(2 \pi)^3}  \, \frac{ \Phi_1 (\Omega, \vec{p}, \theta) \bar{\Phi}_1 (\Omega, \vec{p}, \theta)}{\left[4 m \omega- \vec{k}^2 + i \varepsilon\right]\left[2 m (\omega-\Omega)- (\vec{k} -\vec{p})^2 + i \varepsilon \right]}  \, 
\eeq
We can perform the $\omega$ integration first, as the integrand is sufficiently regular to allow the use of the residue theorem.
The poles of the integrand 
\beq
\omega^{(1)} = \frac{\vec{k}^2}{4 m} - i \varepsilon \, , \qquad
\omega^{(2)} = \Omega + \frac{(\vec{k}- \vec{p})^2}{2 m} - i \varepsilon 
\eeq
are both located in the lower-half complex plane, and so we can close the integration contour in the upper half-plane, obtaining
\beq
\Gamma_1^{(2)} (\Phi_1, \bar{\Phi}_1) = 0 
\eeq
Analogously, in configuration space, 
the vanishing of the two-point function arises from the product of two 
Heaviside functions with opposite arguments, which would have support only in one point. 
By normal ordering, we choose to put this contribution to zero \cite{Bergman:1991hf}.

Since this argument works whenever all the $ \omega $ poles are at the same side of the complex $\omega$ plane (i.e. circulating arrows in the loop), selection rule \ref{srule2} holds in general. However, there is an important caveat: 
the   selection rule relies on the possibility to perform the $\omega$-integration by using the residue theorem,
which in turn requires the integrand to be sufficiently decreasing at infinity for applying Jordan's lemma. 
Propagators should guarantee that this is always the case,
 but $D$-algebra might introduce extra $\omega$ factors
 as a result of the commutation rules \eqref{rules covariant derivatives giving momenta in supergraphs 2}.
 As it will be discussed in \ref{sect-Renormalizability of the theory}, this never happens.
 As a consequence, selection rule \ref{srule2} is true even before performing D-algebra.
 
\begin{figure}[H]
\centering
\begin{subfigure}[b]{0.4\linewidth}
\centering
\includegraphics[scale=0.5]{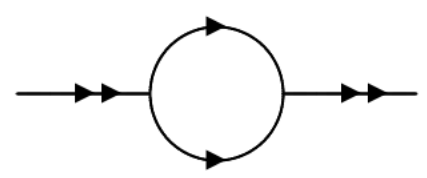}
\caption{}
\end{subfigure} 
\begin{subfigure}[b]{0.4\linewidth}
\centering
\includegraphics[scale=0.25]{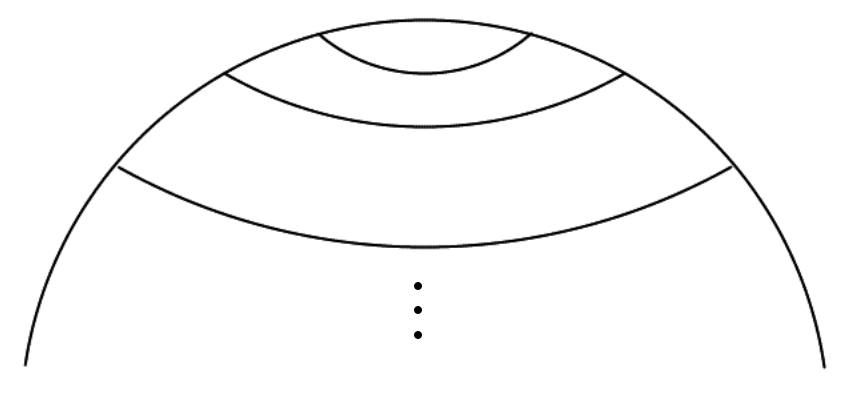}
\caption{}
\end{subfigure}
\caption{Configurations allowed (a) and forbidden (b) by selection rule \ref{srule2}.}
\label{fig6_selectionrule4}
\end{figure}

Selection rule \ref{srule2}  provides fundamental restrictions that make the non-relativistic case very different from the relativistic one, and in this respect easier to study. For example, an immediate consequence is that at one loop, one-particle irreducible diagrams with two external lines admit only one non-vanishing configuration, the one given in fig. \ref{fig6_selectionrule4}(a). 
This rule is true also when the diagram is part of a bigger graph. As a consequence,
the topology shown in fig. \ref{fig6_selectionrule4}(b) is always forbidden, when the number of horizontal lines is bigger than two.
 
\begin{figure}[h]
\centering
\begin{subfigure}[b]{0.4\linewidth}
\centering
\includegraphics[scale=0.4]{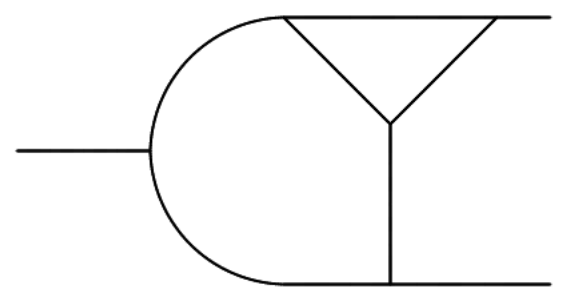}
\caption{}
\end{subfigure}
\begin{subfigure}[b]{0.4\linewidth}
\centering
\includegraphics[scale=0.4]{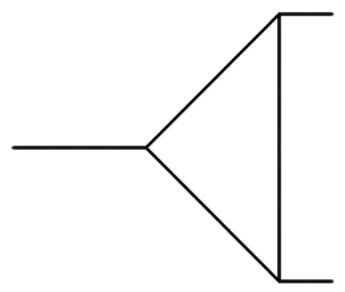}
\caption{}
\end{subfigure} \\
\begin{subfigure}[b]{0.3\linewidth}
\centering
\includegraphics[scale=0.25]{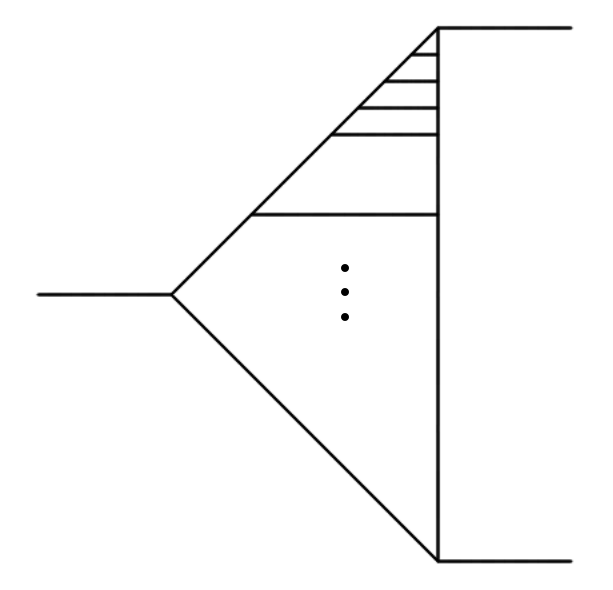}
\caption{}
\end{subfigure}
\begin{subfigure}[b]{0.3\linewidth}
\centering
\includegraphics[scale=0.25]{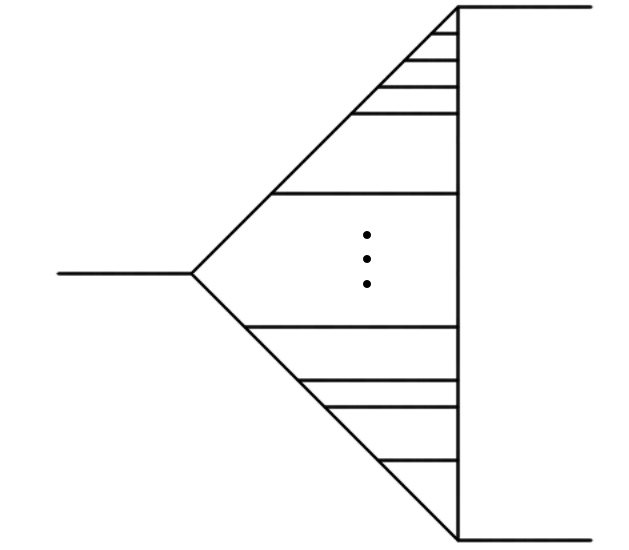}
\caption{}
\end{subfigure}
\begin{subfigure}[b]{0.3\linewidth}
\centering
\includegraphics[scale=0.25]{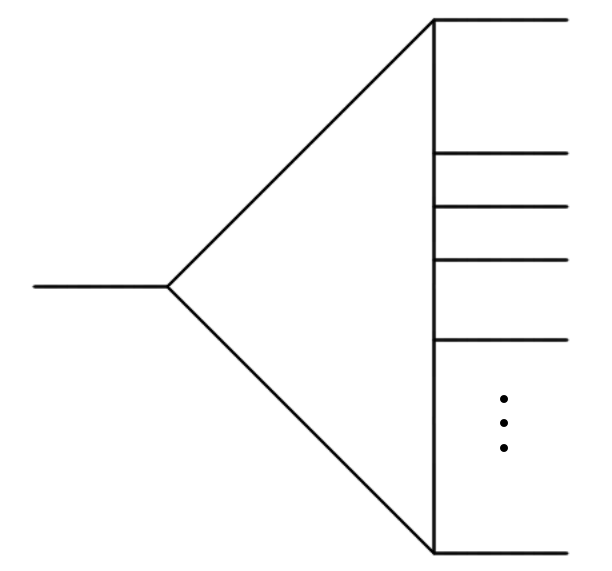}
\caption{}
\end{subfigure}
\caption{Set of vanishing (sub)diagrams due to particle number conservation.
In (e) the number of horizontal lines on the right side is required to be even.}
\label{fig4_selectionrule3}
\end{figure}

Further  selection rules can be obtained from the application of the particle number conservation:
\begin{srule}
\label{srule3}
The (sub)diagrams appearing in fig. \ref{fig4_selectionrule3} are forbidden by particle number conservation.
Configuration {\rm (e)} is forbidden only for an even number of horizontal lines on the right side.
\end{srule}
This statement can be  proved by drawing every possible configuration of arrows 
and checking that no configurations arise, which respect conservation at all vertices.

\begin{figure}[h]
\centering
\begin{subfigure}[b]{0.4\linewidth}
\centering
\includegraphics[scale=0.4]{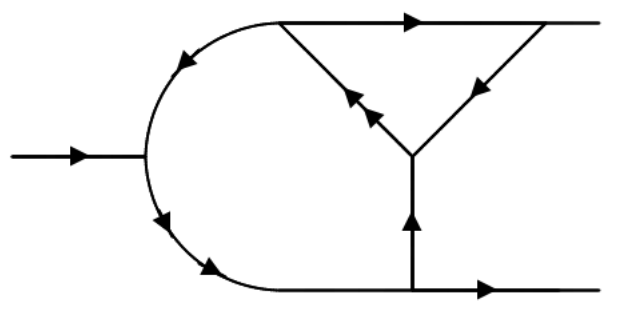}
\caption{}
\end{subfigure}
\begin{subfigure}[b]{0.4\linewidth}
\centering
\includegraphics[scale=0.4]{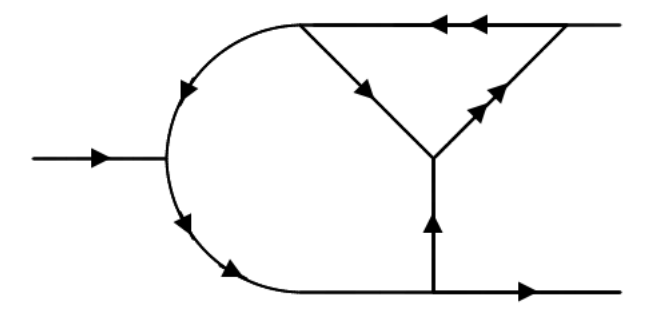}
\caption{}
\end{subfigure}
\begin{subfigure}[b]{0.4\linewidth}
\centering
\includegraphics[scale=0.4]{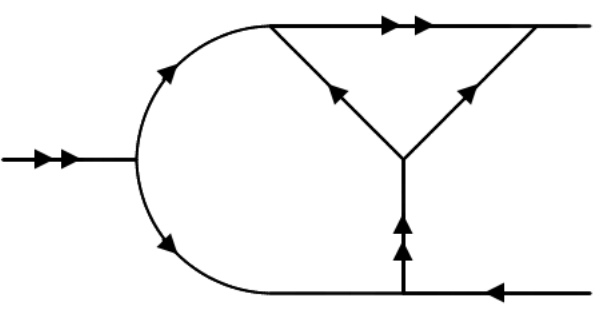}
\caption{}
\end{subfigure}
\begin{subfigure}[b]{0.4\linewidth}
\centering
\includegraphics[scale=0.4]{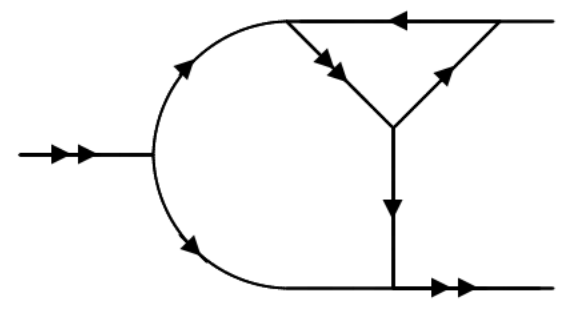}
\caption{}
\end{subfigure}
\begin{subfigure}[b]{0.4\linewidth}
\centering
\includegraphics[scale=0.4]{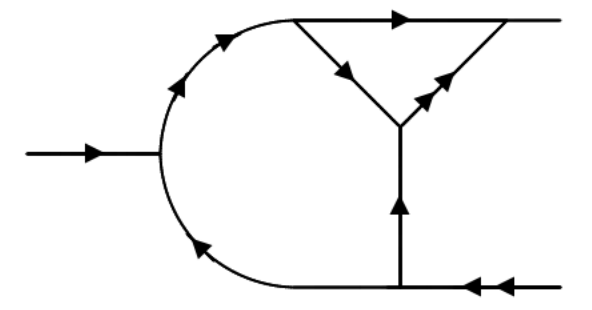}
\caption{}
\end{subfigure}
\begin{subfigure}[b]{0.4\linewidth}
\centering
\includegraphics[scale=0.4]{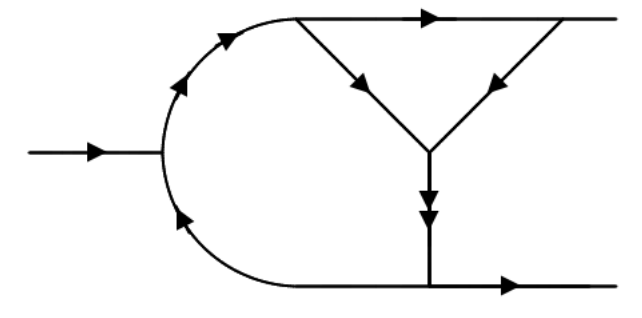}
\caption{}
\end{subfigure}
\caption{All possible configurations of arrows in an example of subdiagram. All of them are forbidden by particle number conservation at the upper right vertex. }
\label{fig5_example_selectionrule3}
\end{figure}

As an example, we consider diagram \ref{fig4_selectionrule3}(a) for which all  possible configurations of arrows are drawn in fig. \ref{fig5_example_selectionrule3}. It can be seen that in all the configurations we cannot consistently assign arrows in the top right vertex.

\subsection{Renormalizability of the theory}
\label{sect-Renormalizability of the theory}

Renormalizability in superspace can be investigated by studying the superficial degree of divergence of a 
generic supergraph with $L$ loops, $E=E_C+E_A$ external lines, 
$P$ internal propagators and $V=V_C+V_A$ vertices, where the $C$  and $A$ subscripts stand for 
chiral and anti-chiral, respectively.  

For a connected graph we can use the topological constraint
\beq
L= P-V+1  
\label{relation between L,P,V}
\eeq
Since there are only cubic vertices, the relation $E + 2P = 3V$ also holds and when combined with the previous constraint leads to 
\beq
P= E + 3 L - 3 
\label{relation with cubic vertices between P,E,L}
\eeq
Finally, in this theory the propagator always connects a chiral vertex with an anti-chiral one. This further restricts the allowed configurations of diagrams, which have to satisfy  
\beq
3 V_A = P + E_A \, , \qquad
3 V_C = P + E_C 
\label{relation on chiral and anti-chiral vertices}
\eeq

The integrand associated to a 1PI supergraph is given by the product of $P$ super-propagators \eqref{superpropagatori momentum space} times a number of $D, \bar{D}$ derivatives acting on the grassmannian delta functions. Counting the number of these derivatives, we have a factor of $ (\bar{D}^2)^2 $ associated to each internal chiral vertex, a factor of $ (D^2)^2$ associated to each internal anti-chiral vertex, a factor of $ \bar{D}^2 $ for each chiral vertex with an external line attached, and finally a factor of $ D^2 $ for each anti-chiral vertex with an external line attached.
The total number of covariant derivatives is then
\beq
(D^2)^{2 V_A-E_A}  (\bar{D}^2)^{2 V_C-E_C} 
\eeq
On the other hand, D-algebra requires one factor $ D^2 \bar{D}^2 $ for each loop to contract the integral to a point in the $ (\theta, \bar{\theta}) $ space. This implies that in non-vanishing diagrams the remaining derivatives   
\beq
(D^2)^{2 V_A-E_A-L}  (\bar{D}^2)^{2 V_C-E_C-L}  
\eeq
are traded with powers of loop momenta, according to the D-algebra procedure explained in Section \ref{sect-Quantum Wess-Zumino model and supergraph formalism}.

Using constraints \eqref{relation on chiral and anti-chiral vertices}, the factor of covariant derivatives associated to the supergraph is given by
\beq
(D^2 \bar{D}^2)^{\frac{2}{3} P - L} \, (D^2)^{- \frac{E_A}{3}} \, (\bar{D}^2)^{-\frac{E_C}{3}}
\eeq

The total contribution from such a diagram is given by this factor multiplied by $P$ propagators $1/\bigtriangleup$ with $ \bigtriangleup \equiv 2M \omega - \vec{k}^2 $, times $L$ integrations on the loop variables.

Looking at the superficial degree of divergence of the integral, the worst case occurs when identities \eqref{rules covariant derivatives giving momenta in supergraphs 2} can be used to trade $ D^2 \bar{D}^2$ with $ \bigtriangleup$, which then cancel internal propagators. The corresponding integral reads
\beq
\hspace{-0.4cm}
 \int d\omega_1 d^2 k_1 \dots d \omega_L d^2 k_L \, \frac{(D^2)^{- \frac{E_A}{3}} \, (\bar{D}^2)^{-\frac{E_C}{3}}}{\bigtriangleup^{L + \frac{P}{3}}} =  \int d\omega_1 d^2 k_1 \dots d \omega_L d^2 k_L  \, \frac{(D^2)^{- \frac{E_A}{3}} \, (\bar{D}^2)^{-\frac{E_C}{3}}}{\bigtriangleup^{2L + \frac{E}{3} -1}} 
\eeq
where in the last step eq. \eqref{relation with cubic vertices between P,E,L} has been used.

Convergence gets even worse in supergraphs where $E_A = E_C = E/2$ 
and the remaining covariant derivatives also combine into inverse propagators. This gives
\beq
  \int d\omega_1 d^2 k_1 \dots d \omega_L d^2 k_L \,  \frac{1}{\bigtriangleup^{2L + \frac{E}{2} -1}} 
\label{counting renormalizability amplitudes}
\eeq
The superficial degree of divergence is $\delta = 2 -E$. It is always negative for $ E \geq 3 $ and the corresponding integrals give finite contributions. 
For self-energy diagrams ($E=2$) logarithmic divergences arise, which can be subtracted by wave-function renormalization. 
Finally, $ E=1$ corresponds to tadpoles, whose prototype is
\beq
\int \frac{d \omega \, d^2 k}{2 M \omega - \vec{k}^2 + i \varepsilon}  
\eeq
After performing the $\omega$-integration, we can use dimensional regularization 
to compute the $\vec{k}$ integral. The result is zero since the integral is 
dimensionful and cannot  depend on any possible scale.
The non-relativistic WZ model is then renormalizable.

However, in the non-relativistic case we have to prove a further property of the amplitudes, i.e. that the loop integrals on $ \omega_1, \dots , \omega_L  $ are separately convergent. This requires the integrand to behave at least as $ 1/\omega^{2}_i $ for a given $\omega_i$-integration. 

To this end we consider a specific loop $L_i$ containing $ P_i $ propagators.
The fact that the inverse of the Galilean propagator is linear in the energy, combined with energy conservation at each vertex implies that the $ P_i $ propagators provide a power $1/ \omega_i^{P_i}$. Since in a loop we always have $P_i \geq 2$ (tadpoles are zero) the convergence of the $\omega_i$-integral is guaranteed, as long as there are no $\omega_i$ powers at numerator. However, D-algebra may generate these powers. In the worst situation $D$-derivatives produce factors which cancel completely some propagators, contracting points in momentum space. In any case, this process leads to a loop with at least two propagators, which is sufficient to ensure the convergence of the integral.
More generally, adjacent loops which in the relativistic case would lead to overlapping divergences, have an even better convergence in $\omega$. 

In conclusion, all the energy integrals are convergent, they do not need to be regularized and we can compute them in the complex plane by using the residue theorem. It is important to stress that this property puts selection rule \ref{srule2} on solid grounds.

\subsection{Loop corrections to the self-energy}
We now study quantum corrections to the WZ model \eqref{non-rel WZ action in superfield formalism}. In order to deal with divergent momentum integrals we use dimensional regularization within the minimal subtraction scheme. Integrals are computed in $d = 2- \epsilon$ and a mass scale $\mu$ is introduced to rescale dimensionful quantities. We define renormalized quantities \beq
\begin{cases}\label{ren_functions}
\Phi_a = Z_a^{-1/2} \,  \Phi_a^{(B)} = \le 1- \frac12 \delta_a \ri \Phi_a^{(B)} \qquad a=1,2\\
g = \mu ^{-\epsilon} Z_g^{-1} g^{(B)} = \mu^{-\epsilon}  (1 - \delta_g) g^{(B)} 
\end{cases}
\eeq
and determine counterterms proportional to $\delta_a$, $\delta_g$ 
\beq
\mathcal{L}_{\mathrm{ren}}   +
  \int d^4 \theta \, \le \delta_1 \bar{\Phi}_1 \Phi_1 + \delta_2 \bar{\Phi}_2 \Phi_2  \ri  + \int d^2 \theta \, \left[
\mu^{\epsilon} g \le \delta_g + \delta_1 + \frac12 \delta_2 \ri  \Phi_1^2 \Phi_2 \right] + \mathrm{h.c.} 
\eeq
which cancel UV divergences.  

\subsubsection*{One loop}
\begin{figure}[h]
\centering
\begin{subfigure}[b]{0.45\linewidth}
\includegraphics[scale=1.2]{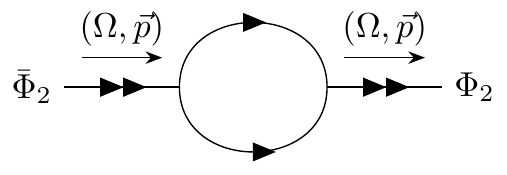}
\end{subfigure}
\begin{subfigure}[b]{0.45\linewidth}
\end{subfigure}
\caption{One-loop correction to the self-energy of $\Phi_2$.}
\label{fig7_rinormalizzione self-energy supercampo 2}
\end{figure}

By applying  selection rule \ref{srule2} to sector 1, we immediately realize that there are no allowed one-loop diagrams, except for 
 the one depicted in fig. \ref{fig3_rinormalizzione self-energy supercampo 1}, which was shown to vanish.
As a consequence,
\beq
\delta_1^{\rm (1loop)} = 0  
\label{wavefunction renormalization sector 1}
\eeq
For the one-loop self-energy in sector 2 we find, instead, the allowed diagram in fig. \ref{fig7_rinormalizzione self-energy supercampo 2}.

After performing D-algebra, the corresponding contribution reads
\beq
i \Gamma_2^{(2)} (\Phi_2, \bar{\Phi}_2) =  2 |g|^2  \int d^4 \theta \,   \frac{d\omega \, d^2 k}{(2 \pi)^3} \,  \frac{ \Phi_2 (\Omega, \vec{p}, \theta) \bar{\Phi}_2 (\Omega, \vec{p}, \theta) }{\left[2 m \omega- \vec{k}^2 + i \varepsilon \right] \left[ 2 m (\Omega-\omega)- (\vec{p} -\vec{k})^2 + i \varepsilon \right]}   
\eeq
As already discussed, the $ \omega $-integral is convergent and can be easily performed by means of the residue theorem, leading to
\beq
 \Gamma_2^{(2)} (\Phi_2, \bar{\Phi}_2) = - \frac{|g|^2 }{m}  \int d^4 \theta  \, \Phi_2 (\Omega, \vec{p}, \theta) \bar{\Phi}_2 (\Omega, \vec{p}, \theta) \,  \int \frac{d^2 k}{(2 \pi)^2} \, \frac{1}{2 m \Omega - \vec{k}^2 - ( \vec{p} - \vec{k})^2 + i \varepsilon}   
\eeq
The two-dimensional momentum integral can be now performed using dimensional regularization. Focusing on its divergent part we obtain
\beq
\Gamma_2^{(2)} (\Phi_2, \bar{\Phi}_2) \to  \frac{|g|^2}{4 \pi m} \frac{1}{\varepsilon} \int d^4 \theta \, \Phi
_2 (\Omega, \vec{p}, \theta) \bar{\Phi}_2 (\Omega, \vec{p}, \theta)  
\eeq
In the minimal subtraction scheme this leads to the following counterterm  
\beq\label{superfield_ren}
\delta_2^{\rm (1loop)} = - \frac{|g|^2}{4 \pi m} \frac{1}{\varepsilon} 
\eeq

\subsubsection*{Two loops}

Selection rules \ref{srule2} and \ref{srule3} are sufficient to rule out any two-loop correction to the self-energies. In fact, looking at the two possible two-loop topologies of diagrams depicted in fig. \ref{fig8_selfenergy a 2 loop}, it is easy to realize that no consistent assignments of arrows exist, or they vanish due to circulating arrows in a loop.

\begin{figure}[h]
\centering
\begin{subfigure}[b]{0.4\linewidth}
\includegraphics[scale=0.3]{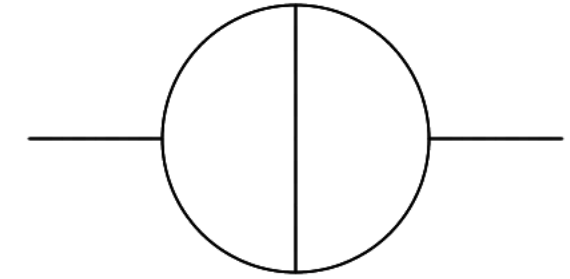}
\end{subfigure}
\begin{subfigure}[b]{0.4\linewidth}
\includegraphics[scale=0.3]{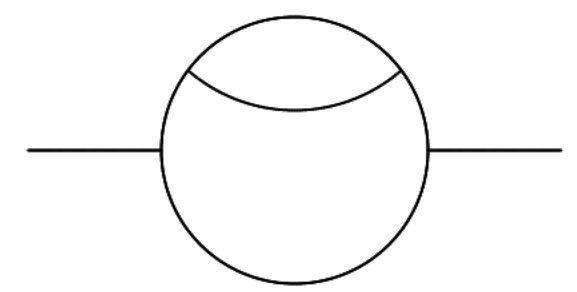}
\end{subfigure}
\caption{Topologies of possible two-loop corrections to the self-energies.}
\label{fig8_selfenergy a 2 loop}
\end{figure}

\subsubsection*{Three loops}
At three loops, the set of diagrams is of course richer. In the relativistic setting all possible diagrams have been given in \cite{Abbott:1980jk}, where the three-loop $ \beta $-function was computed.

In the non-relativistic case selection rules \ref{srule2} and \ref{srule3} discard almost all possible configurations, leading only to one non-trivial type of diagram, the non-planar one depicted in fig. \ref{fig8_selfenergy a 2 loop}.
However, looking at all possible assignments of arrows we conclude that there is always a circulating loop, which entails a vanishing result according to selection rule \ref{srule2}. Therefore, there are no three-loop corrections to the self-energies of both superfields.

\begin{figure}[h]
\centering
\begin{subfigure}[b]{0.45\linewidth}
\centering
\includegraphics[scale=0.2]{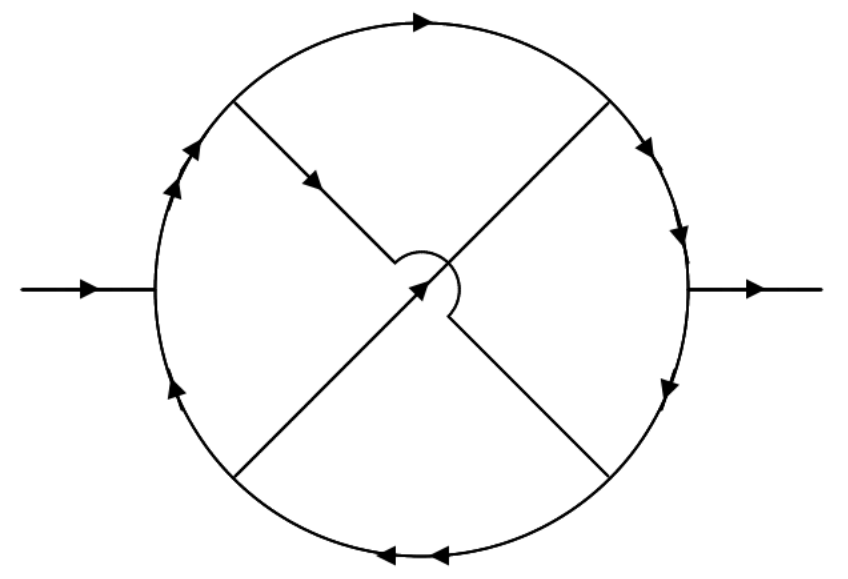}
\end{subfigure}
\begin{subfigure}[b]{0.45\linewidth}
\centering
\includegraphics[scale=0.2]{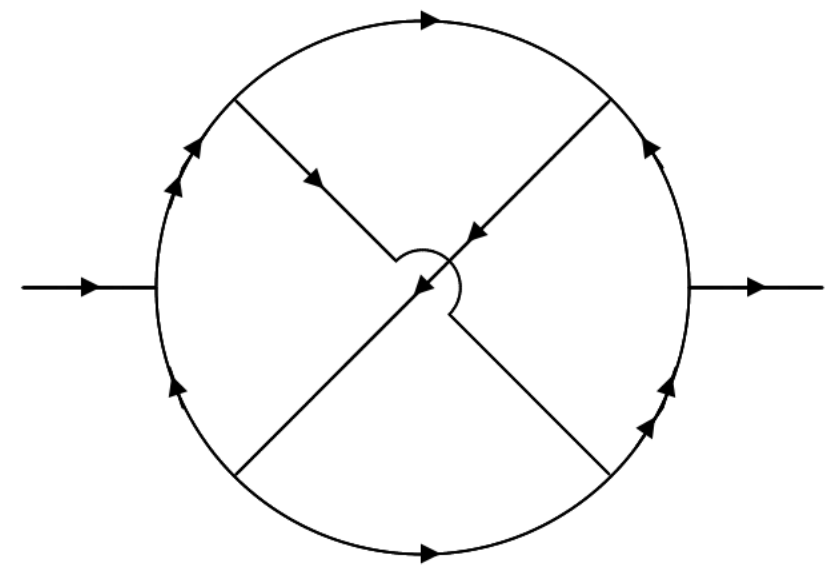}
\end{subfigure}
\begin{subfigure}[b]{0.45\linewidth}
\centering
\includegraphics[scale=0.26]{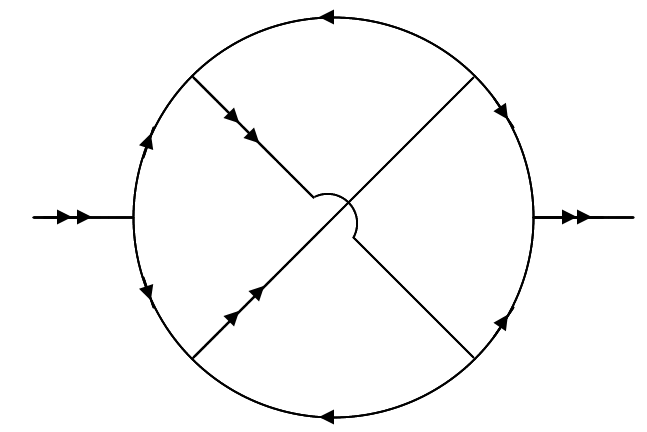}
\end{subfigure}
\begin{subfigure}[b]{0.45\linewidth}
\centering
\includegraphics[scale=0.2]{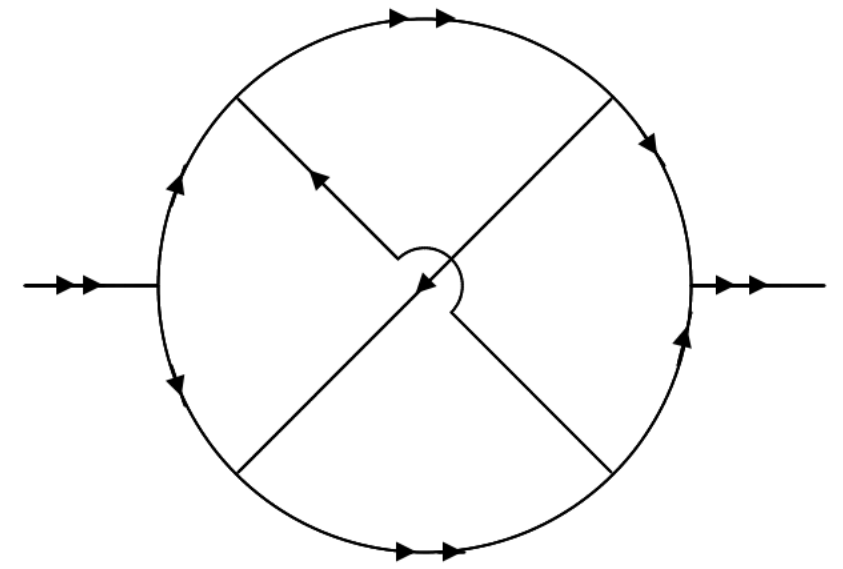}
\end{subfigure}
\caption{Non-trivial three-loop corrections to the self-energies. We depict all possible assignments of arrows in the lines.}
\label{fig8_selfenergy a 2 loop}
\end{figure}

\subsubsection*{Four loops}
At four loops all possible configurations of self-energy supergraphs can be imported from \cite{Sen:1981hk}, where the $\beta $-function for the relativistic WZ model was computed. 

\begin{figure}[h]
\centering
\begin{subfigure}[b]{0.32\linewidth}
\centering
\includegraphics[scale=0.18]{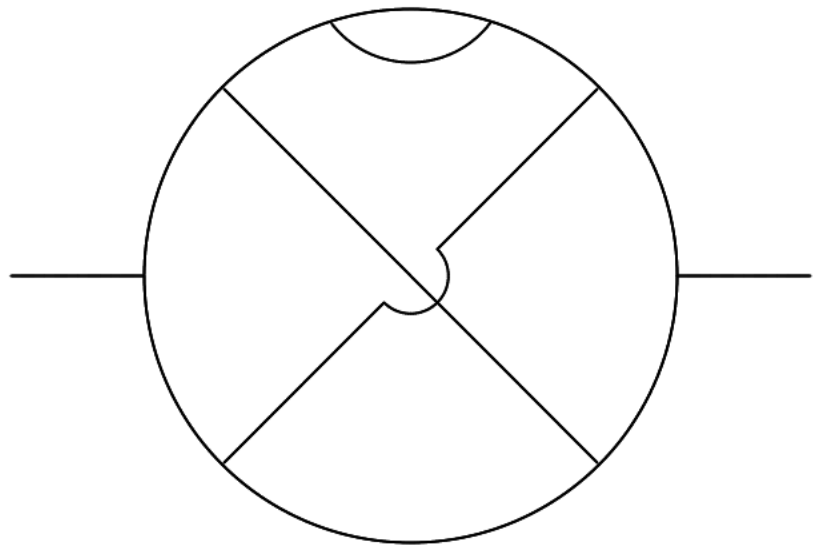}
\subcaption{}
\end{subfigure}
\begin{subfigure}[b]{0.32\linewidth}
\centering
\includegraphics[scale=0.18]{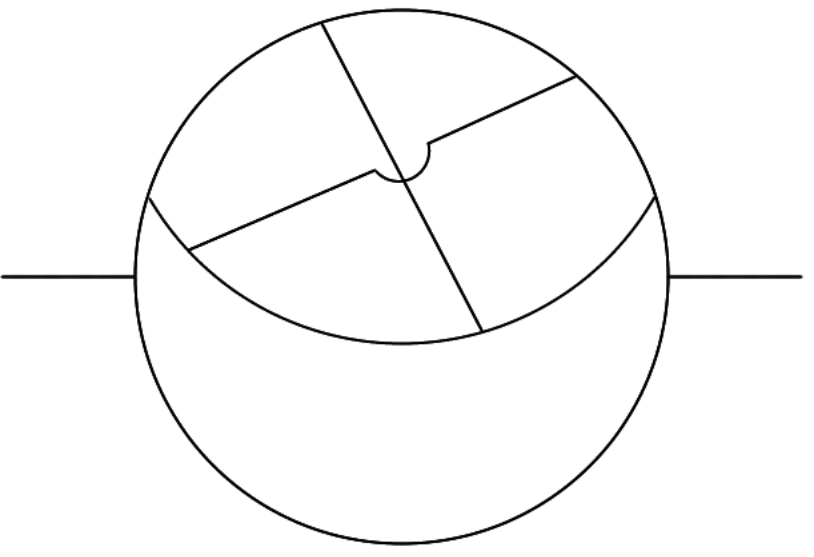}
\subcaption{}
\end{subfigure}
\begin{subfigure}[b]{0.32\linewidth}
\centering
\includegraphics[scale=0.18]{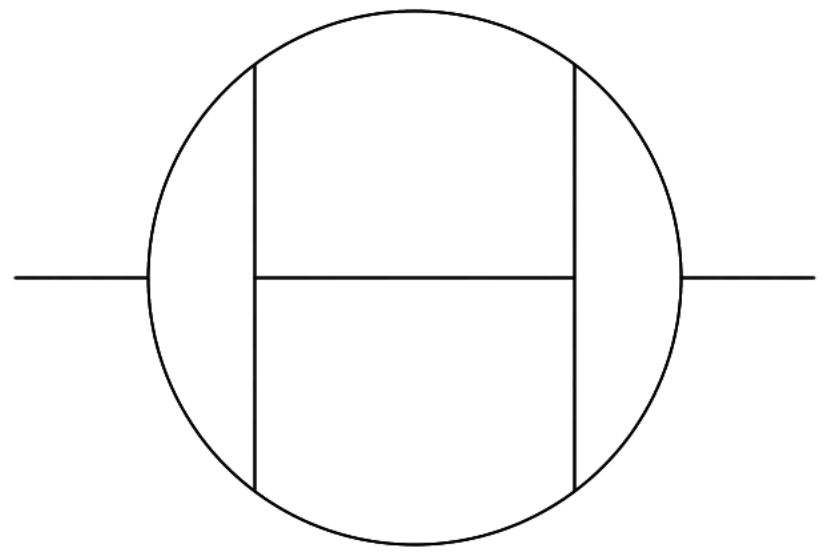}
\subcaption{}
\end{subfigure}
\caption{Non-trivial quantum corrections to the self-energy at four loops.}
\label{fig9_selfenergy_4_loops}
\end{figure}

In the non-relativistic case, once again, the selection rules discard almost all configurations, leaving only the non-trivial diagrams listed in fig. \ref{fig9_selfenergy_4_loops}(a)-(c). 

\begin{figure}[h]
\centering
\begin{subfigure}[b]{0.35\linewidth}
\centering
\includegraphics[scale=0.18]{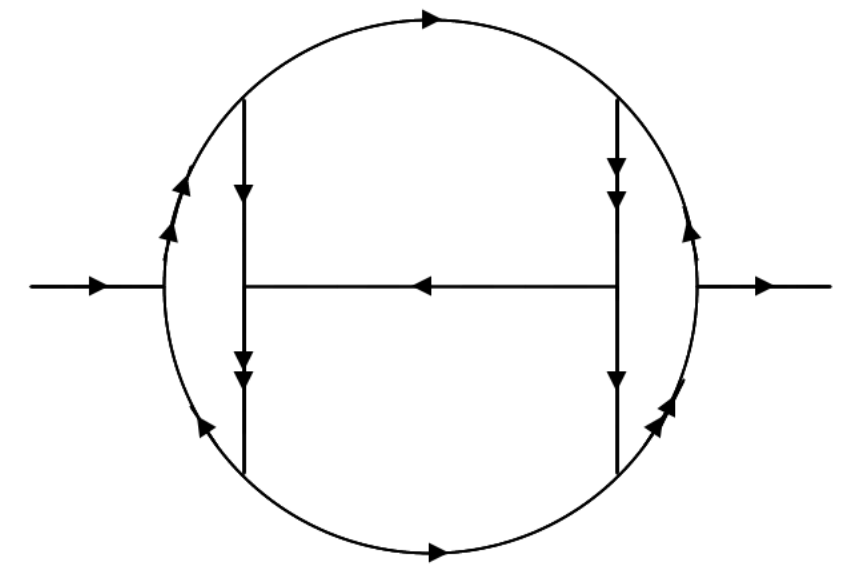}
\subcaption{}
\end{subfigure}
\begin{subfigure}[b]{0.35\linewidth}
\centering
\includegraphics[scale=0.18]{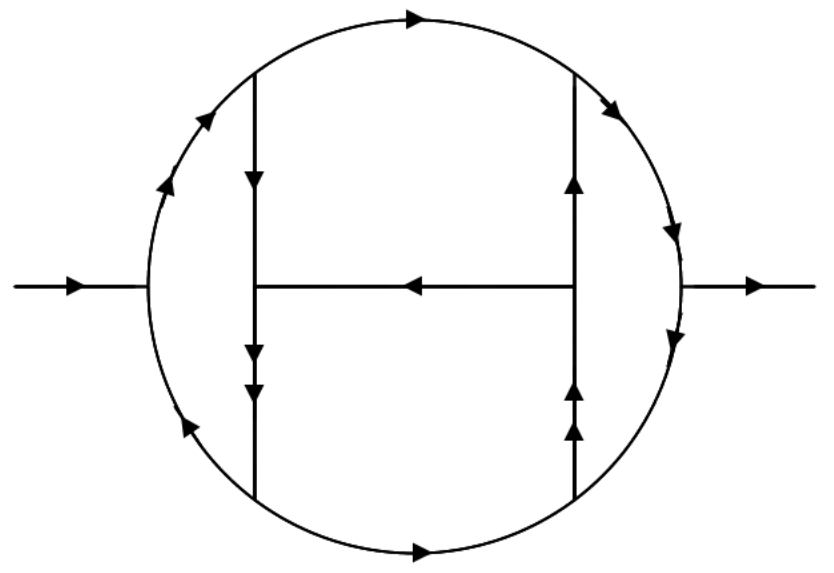}
\subcaption{}
\end{subfigure}
\begin{subfigure}[b]{0.35\linewidth}
\centering
\includegraphics[scale=0.18]{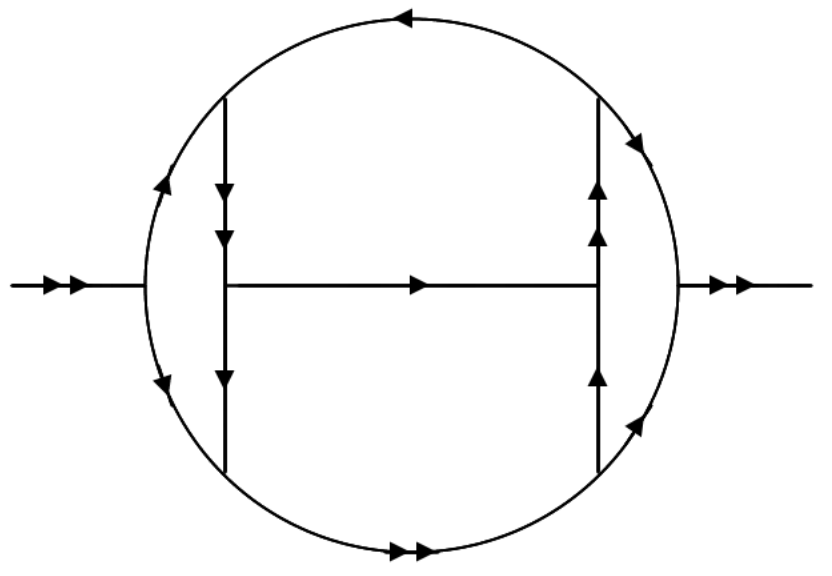}
\subcaption{}
\end{subfigure}
\begin{subfigure}[b]{0.35\linewidth}
\centering
\includegraphics[scale=0.18]{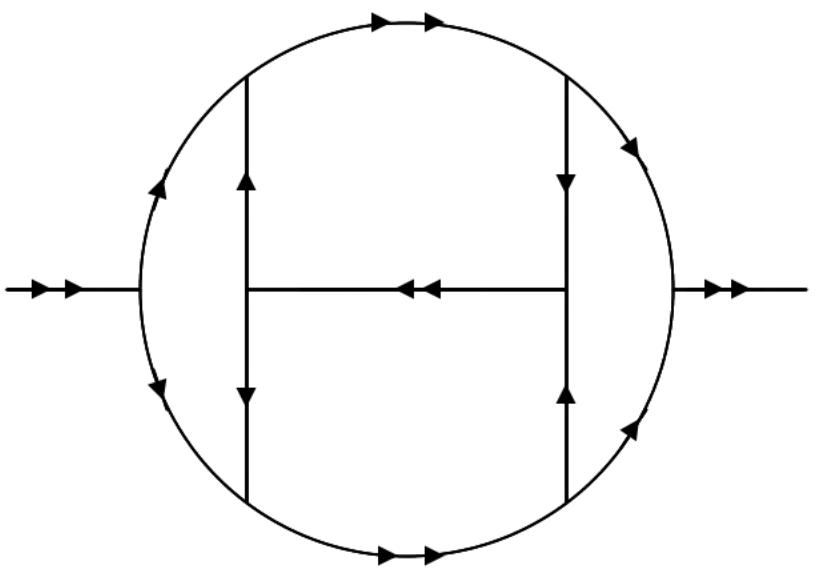}
\subcaption{}
\end{subfigure}
\caption{Allowed assignments of arrows to the lines of  diagram (c) in Fig. \ref{fig9_selfenergy_4_loops}.} 
\label{fig10_examples_selfenergy_4_loops}
\end{figure}

The first two graphs contain as a subgraph the non-planar three-loop diagram already discussed. Therefore, with similar arguments, we can prove that these diagrams vanish. Diagram (c) is a new configuration and in principle it allows for different configurations of arrows depicted in fig. \ref{fig10_examples_selfenergy_4_loops}(a)-(d).
It is easy to see that all of them contain circulating loop arrows, thus they vanish by means of selection rule \ref{srule2}.


\subsubsection*{Higher loops}
Up to four loops we have found that non-vanishing quantum corrections to the self-energy appear only in sector 2 and only at one loop. 
Triggered by these results, the natural question which arises is whether the same pattern repeats at any loop order or we should expect non-vanishing contributions at higher loops. 

\begin{figure}[h]
\centering
\begin{subfigure}[b]{0.4\linewidth}
\centering
\includegraphics[scale=0.18]{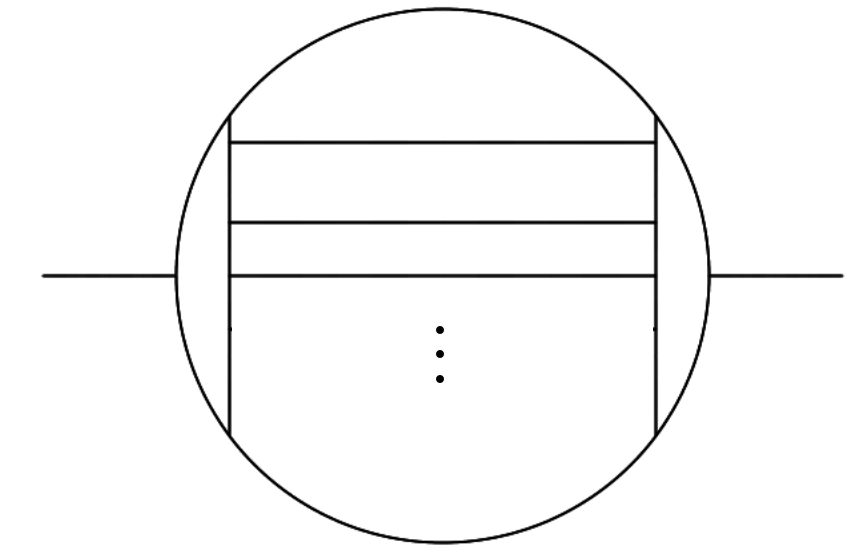}
\subcaption{}
\end{subfigure}
\begin{subfigure}[b]{0.4\linewidth}
\centering
\includegraphics[scale=0.18]{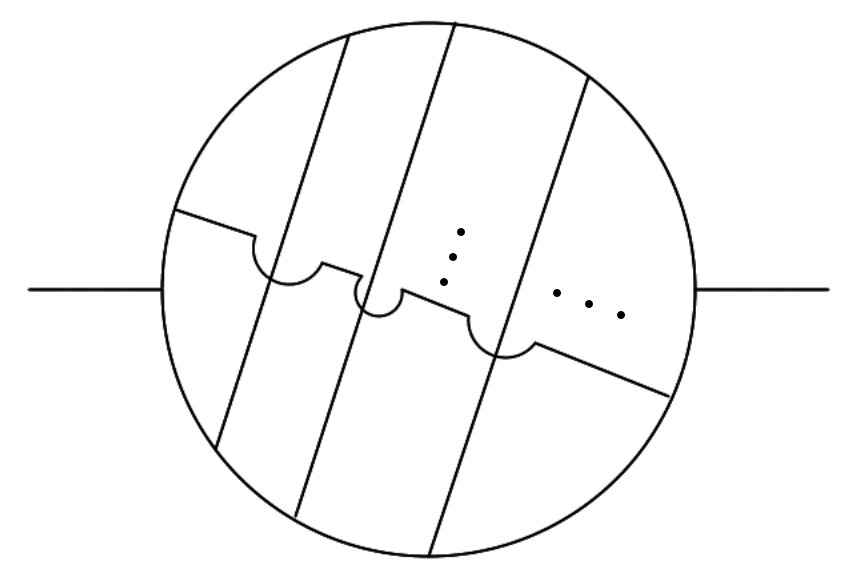}
\subcaption{}
\end{subfigure}
\begin{subfigure}[b]{0.4\linewidth}
\centering
\includegraphics[scale=0.18]{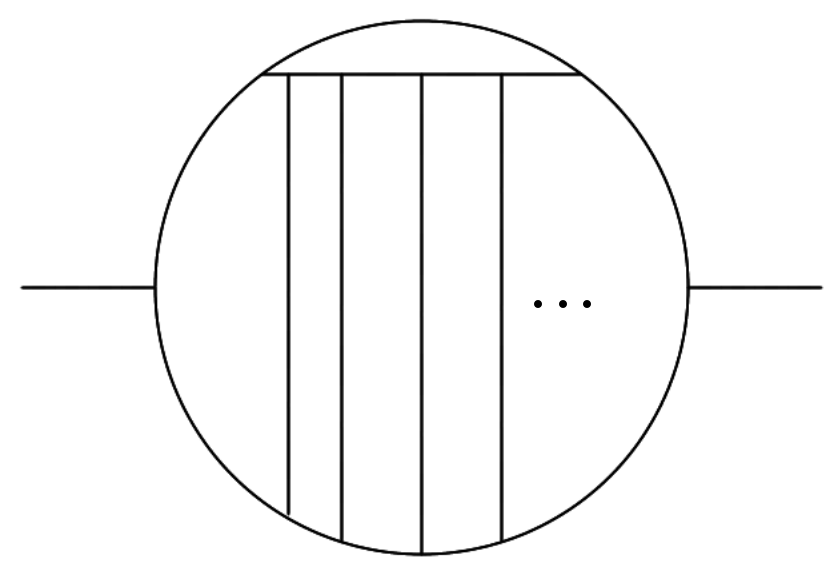}
\subcaption{}
\end{subfigure}
\begin{subfigure}[b]{0.4\linewidth}
\centering
\includegraphics[scale=0.18]{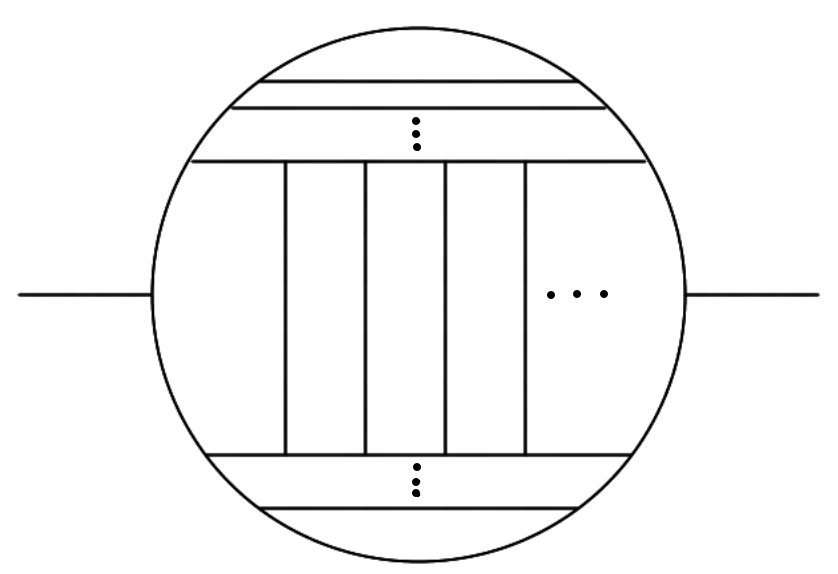}
\subcaption{}
\end{subfigure}
\begin{subfigure}[b]{0.4\linewidth}
\centering
\includegraphics[scale=0.18]{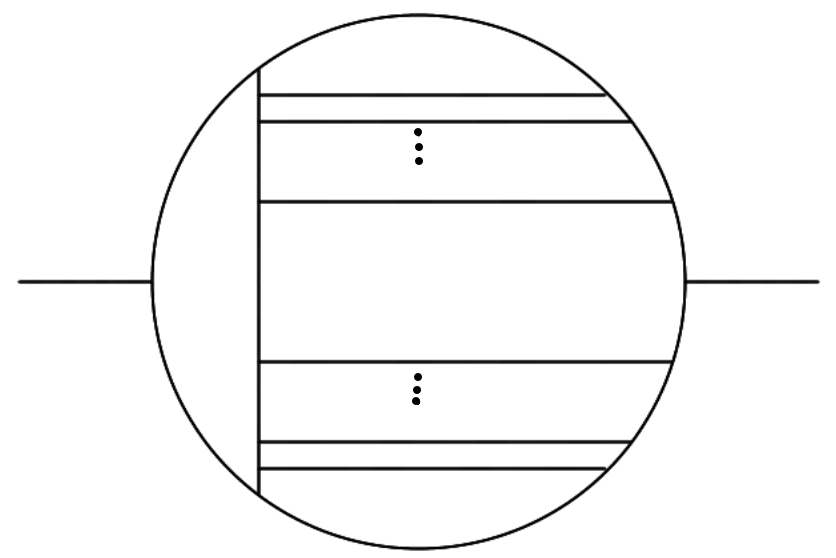}
\subcaption{}
\end{subfigure}
\caption{Non-trivial vanishing quantum corrections to self-energies at generic loop level.}
\label{fig11_candidates_nloop_prop}
\end{figure}

To face this question, we can start excluding a large class of diagrams which contain recursive structures that can be shown to vanish due to the selection rules. First, all diagrams containing the structures  in fig. \ref{fig4_selectionrule3} are forbidden by selection rule \ref{srule3}. Among the allowed ones we can select for instance the structures  in fig. \ref{fig11_candidates_nloop_prop}. 
Generalizing the previous analysis one can check that all these structures always contain a closed loop of arrows and are eventually ruled out by selection rule \ref{srule2}. This remains true for all the diagrams that can be obtained by gluing different structures of fig. \ref{fig11_candidates_nloop_prop}. 

\begin{figure}[h]
\centering
\begin{subfigure}[b]{0.4\linewidth}
\centering
\includegraphics[scale=0.18]{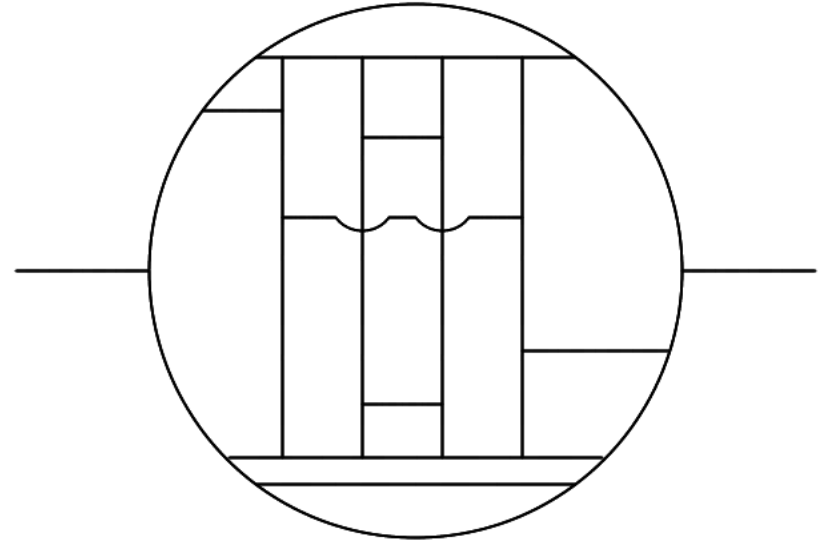}
\end{subfigure}
\caption{General self-energy diagram.}
\label{general_diagram}
\end{figure}

Although these topologies cover a vast number of diagrams, they are not exhaustive and in principle we cannot exclude the appearance of possible non-vanishing contributions from more general configurations, like the one in fig. \ref{general_diagram}. Nonetheless, based on the experience gained up to four loops, we expect that when the numbers of loops increases it becomes more and more difficult to realize configurations of arrows without closed loops. Therefore, we can quite safely conjecture that the self-energy of the $\Phi_1$ superfield is not corrected at quantum level, while the one for $\Phi_2$ is one-loop exact.

Independently of the validity of this conjecture, it is a matter of fact that in the non-relativistic model self-energy corrections are much simpler than their 4d ${\cal N}=1$ relativistic cousins. In fact, in the relativistic case the kinetic term acquires UV divergent corrections at any loop order. Instead, in the non-relativistic model the particle number conservation and the structure of the propagators drastically reduce the number of non-vanishing contributions, leading to a theory which is one-loop exact. In particular, this shows that at quantum level the non-relativistic three-dimensional ${\cal N}=2$ WZ model cannot be obtained simply from null reduction of the four-dimensional relativistic model.

\subsection{Loop corrections to the vertices}

As already discussed in section \ref{sect-Renormalizability of the theory} for a number of external legs equal or greater than three we do not expect UV divergent contributions. Moreover, given that the null reduction does not affect the spinorial part of superspace, we do not expect chiral integrals to be produced. Therefore, the perturbartive non-renormalization theorem for the superpotential should still work. As a consequence, the following constraint on the renormalization functions \eqref{ren_functions} follows 
\beq\label{ren_constraint}
\delta_g + \delta_1 + \frac12 \delta_2 = 0 \quad \Rightarrow \quad \delta_g^{\rm (1loop)} = \frac{|g|^2}{8\pi m } \frac{1}{\epsilon}
\eeq

It is worth discussing the three-vertex diagrams anyway in order to investigate how the selection rules restrict the number of possible quantum corrections for configurations with three external fields.

As in the relativistic case, at one-loop there is no way to draw any three-point diagram as long as the model is massless. 

\begin{figure}[h]
\centering
\begin{subfigure}[b]{0.45\linewidth}
\centering
\includegraphics[scale=0.18]{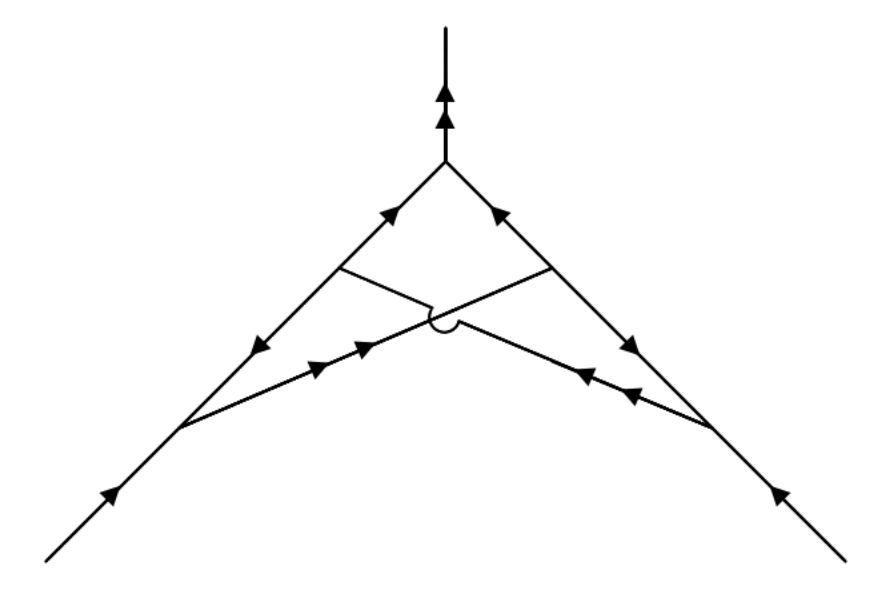}
\caption{}
\end{subfigure}
\begin{subfigure}[b]{0.45\linewidth}
\centering
\includegraphics[scale=0.18]{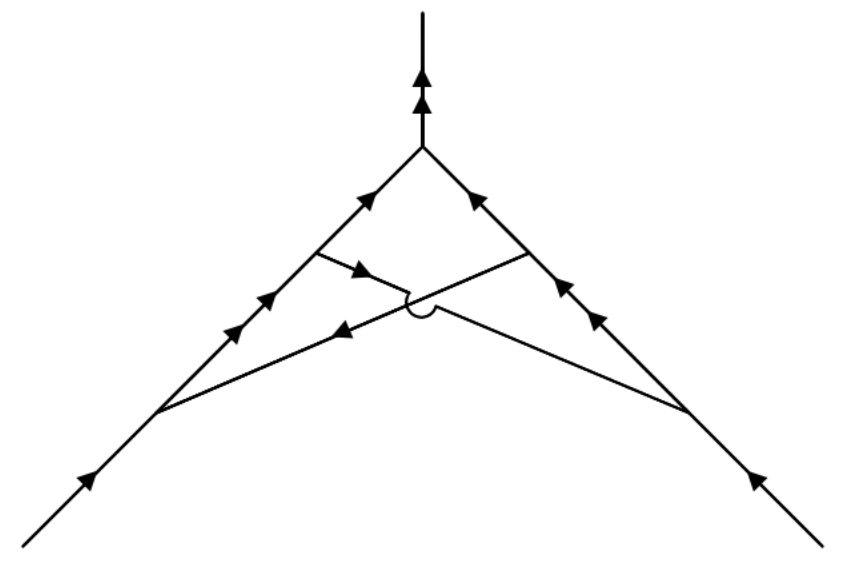}
\caption{}
\end{subfigure}
\begin{subfigure}[b]{0.45\linewidth}
\centering
\includegraphics[scale=0.18]{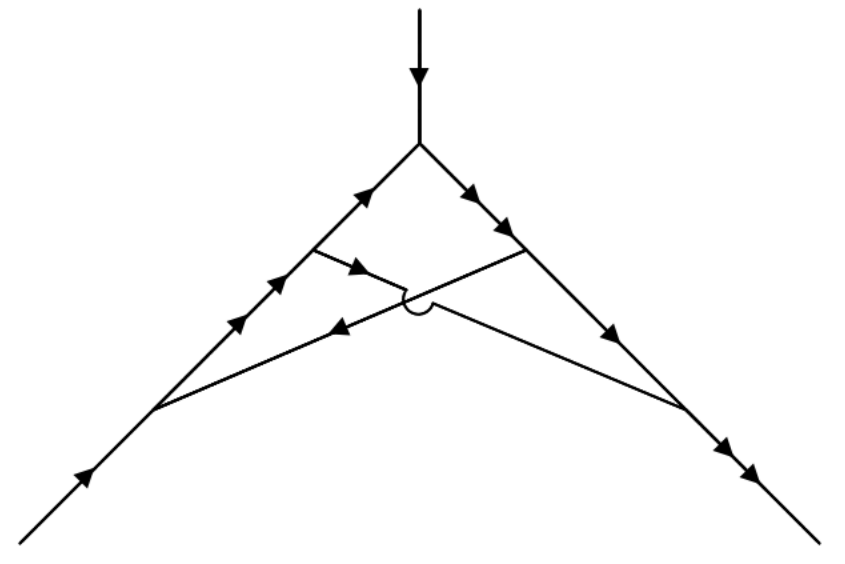}
\caption{}
\end{subfigure}
\begin{subfigure}[b]{0.45\linewidth}
\centering
\includegraphics[scale=0.18]{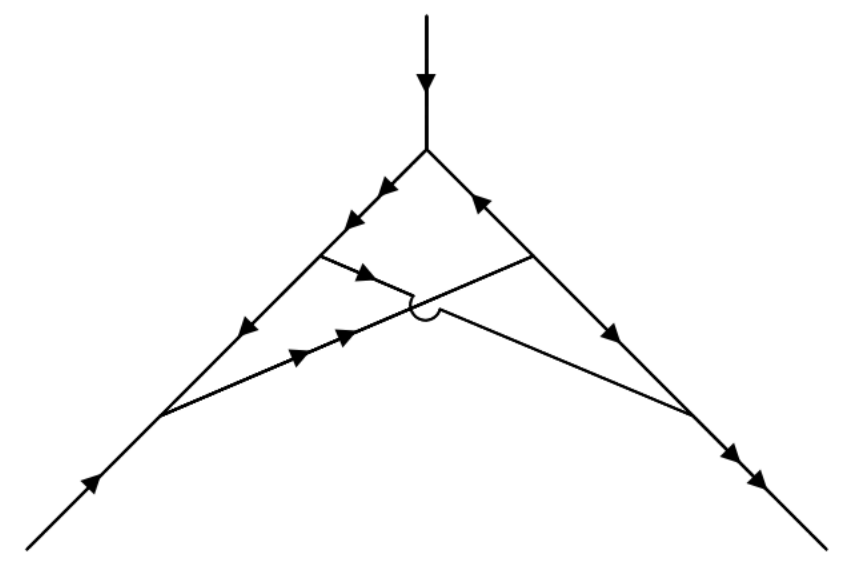}
\caption{}
\end{subfigure} 
\caption{Two-loop 1PI diagram for the three-point vertex. We depicted all the configurations of arrows associated to the lines.}
\label{fig14_2 loop supervertici a 3}
\end{figure}

At two loops the only supergraph allowed by particle number conservation is the one in fig. \ref{fig14_2 loop supervertici a 3}, where all possible configurations of arrows have been depicted. In all the diagrams we see that a circulating loop of arrows appears, thus this diagram is ruled out by selection rule \ref{srule2}. In appendix \ref{examples} we present details of the calculation. 

Extending this analysis at higher loops, again we find that in the non-relativistic model the number of (finite) quantum contributions is drastically reduced compared to the relativistic case.

\subsection{Non-relativistic non-renormalization theorem}

Strong support to the conjectured absence of corrections to the three-point vertex comes from the existence of a non-relativistic version of the non-renormalization theorem. Here we argue that such a theorem can be easily inherited from the relativistic one. 

We consider a generic Galilean WZ model for $n$ chiral superfields
\beq
S= \int d^3 x d^2 \theta d^2 \bar{\theta} \, \bar{\Phi}_a \Phi_a + g \int d^3 x d^2 \theta \, W(\Phi_a) + \mathrm{h.c.}
\label{WZ-generico-NR}
\eeq
obtained by null reduction of the relativistic one in eq. (\ref{WZ-generico}).

The same argument used in the relativistic case can be adapted here in order to
rule out quantum corrections to the F-term.  As in eq. (\ref{WZ-generico2})
we can introduce one extra chiral superfield $Y$ multiplying the superpotential,
which can be set equal to $1$ in order to reproduce eq. (\ref{WZ-generico-NR}), together with wave function renormalization superfields $Z_{ab}$
\beq
\tilde{S}= \int d^3 x \, d^4 \theta  \, Z_{ab} \bar{\Phi}_a \Phi_b + \int d^3 x \, d^2 \theta \,  
Y \, W(\Phi_a) + \mathrm{h.c.} 
\eeq
Since the non-relativistic limit via null reduction technique does not modify the grassmannian part of the superfields in the action,
R-symmetry works in the same way as in the relativistic case. Therefore, as in the relativistic case, we assign R-charges  $R(\Phi_a)=0$
and  $R(Y)=2$. 

The regularization that we used, which corresponds to first performing the regular $\omega$-integrals and then the $\vec{k}$-integrals in dimensional regularization, 
preserves SUSY. Therefore, the Wilsonian effective action at a given scale $\lambda$ will have the following general structure
\beq
\tilde{S}_\lambda= \int d^3 x \, d^4 \theta   \,  K(\bar{\Phi}_a \Phi_a,Z_{ab}, Y, \bar{Y}, D)
+\int d^3 x \, d^2 \theta \, W_\lambda (\Phi_a, Y) 
\eeq
Now, R-invariance and holomorphicity of the superpotential, combined with the weak coupling limit, give as in the relativistic case
 $W_\lambda=Y \,  W(\Phi_a)$.

\section{Conclusions}
\label{Conclusions}

In this paper we studied the renormalization properties of a $2+1$ dimensional model
with $\mathcal{S}_2 \mathcal{G}$ Galilean SUSY invariance, 
obtained by null reduction of the four-dimensional relativistic Wess-Zumino model
(see eq. (\ref{WZM-before-null-red})). 
The model contains two chiral superfields and it is the simplest  theory
with F-term interactions compatible with 
the non-relativistic $U(1)$ mass invariance. 
When reduced in components, after integrating out the auxiliary fields, the interaction lagrangian includes
cubic derivative and quartic non-derivative local interactions.

We  constructed the ${\cal N}=2$ non-relativistic superspace as null reduction of the four-dimensional ${\cal N}=1$ relativistic one, 
and quantized the system in this framework. 
Since null reduction does not affect the grassmanian coordinates and SUSY is totally preserved, 
loop corrections to the cubic vertex obey a non-relativistic
analog of the non-renormalization theorem for the F-terms of the parent
relativistic theory. 

More interestingly, in the non-relativistic case, $U(1)$ invariance and the retarded nature 
of the field propagators lead to further selection rules that forbid many topologies of potentially divergent diagrams.  As a result,
up to four loops, quantum corrections to the 
K\"ahler  potential are all vanishing with the exception
of the one-loop contribution to the $\Phi_2$ propagator.

Extending the investigation to higher loops we have provided strong evidence that the combination of the non-renormalization properties of the F-terms and the selection rules, in particular the vanishing
of loop diagrams whose arrows form a closed loop, suppresses Galilean UV divergences in a very efficient way and makes the model one-loop exact.  
Therefore, the non-relativistic model has peculiar quantum properties that render it very different from its relativistic counterpart. 
At the quantum level, it cannot be simply obtained 
by null reduction of the ordinary WZ model in $3+1$ dimensions. 

The result we have found is reminiscent of relativistic gauge theories with extended SUSY, like for instance the relativistic $ \mathcal{N}=2 $ SYM in 3+1 dimensions. 
In that case extended supersymmetry constrains the corrections to the K\"ahler  potential to be related to the F-terms, which are protected by the non-renormalization theorem.
In the non-relativistic model discussed in this paper, instead the protection of the K\"ahler potential is related to the $U(1)$ charge conservation at each vertex, which in many
diagrammatic contributions constrains arrows to form a closed loop, so leading to a vanishing integral.
It would be interesting to investigate 
if a common hidden mechanism exists, which is responsible of the similar
mild UV behavior of these two rather different classes of theories.

At the classical level the model is scale-invariant. However, this symmetry is broken by anomalous quantum corrections. As follows from our result \eqref{ren_constraint}
the beta function for the coupling $g$ is given by
\beq
\beta_g=\frac{d g } {d \log \mu} =\frac{g^3}{ 4 \pi m}
\label{betafunction}
\eeq
and the theory is infrared free at low energies, 
like the model studied in \cite{Bergman:1991hf}.
If the  K\"ahler potential  is one-loop exact,
as it seems to be the case, then equation (\ref{betafunction}) is exact.
Note that in the single field example without SUSY studied 
in  \cite{Bergman:1991hf} the beta function is also one-loop exact,
but for different reasons. There the propagator has no quantum corrections
and all the UV divergences come from the vertex corrections.

As a continuation of the study of non-relativistic conformal anomalies
in curved Newton-Cartan (NC) background
\cite{Jensen:2014hqa,Arav:2016xjc,Auzzi:2015fgg,Auzzi:2016lxb,Pal:2017ntk,Pal:2016rpz,Auzzi:2016lrq},
it would be interesting to study superconformal anomalies
of a Galilean SUSY theory in the presence of a classical NC supergravity source.
 We leave this as a topic for further work.

A non-relativistic theory of a chiral superfield coupled 
to a Chern-Simons gauge field, which   is invariant under the 
conformal extension of the $ \mathcal{S}_2 \mathcal{G}$
algebra, was constructed in \cite{Leblanc:1992wu}.
We expect that further examples  of
 $ \mathcal{S}_2 \mathcal{G}$ theories may be constructed by coupling 
 the F-term interacting theory discussed in this paper to a Chern-Simons gauge field.
 These examples should contain trilinear derivative couplings
 between scalars and fermions and then
 should be different from existing constructions of non-relativistic
 SUSY Chern Simons theory built from the $c \rightarrow \infty$
 non relativistic limit (see e.g. \cite{Nakayama:2008qz,Nakayama:2008td,Lee:2009mm,Lopez-Arcos:2015cqa,Doroud:2015fsz}).
 In analogy to the non-SUSY example studied in \cite{Bergman:1993kq},
 we expect that for special values of the F-term coupling $g$ the
 resulting theory is conformal.   We leave a detailed study as
 a topic for further investigation.
 These extensions may provide a useful theoretical SUSY setting  
 for studying  non-abelian anyons \cite{Doroud:2015fsz,Doroud:2016mfv}
 and non-relativistic particle-vortex dualities \cite{Turner:2017rki}.

\vskip 25pt

\acknowledgments

This work has been supported in part by Italian Ministero dell'Istruzione, Universit\`a e Ricerca (MIUR), and by Istituto
Nazionale di Fisica Nucleare (INFN) through the  ``Gauge and String Theory'' (GAST) and ``Gauge Theories, Strings, Supergravity'' (GSS)   research projects.

\newpage

\addtocontents{toc}{\protect\setcounter{tocdepth}{1}}
\appendix

\section{Conventions}
\label{app-conv}

In this Appendix we collect SUSY conventions in $3+1$ and $2+1$ dimensions. For conventions in four dimensions we primarily refer to \cite{Martin:1997ns}.

\subsection*{Spinors}
In $3+1$ dimensional Minkowski space-time we take the metric 
$\eta^{MN} = \mathrm{diag} (-1,1, 1 , 1) $
and denote left-handed Weyl spinors as $ \psi_{\alpha}$, while right-handed ones as $ \bar{\chi}^{\dot{\alpha}}$.

Spinorial indices are raised and lowered as
\beq\label{raise&lower}
\psi^{\alpha} = \epsilon^{\alpha \beta} \psi_{\beta} \, , \qquad
\bar{\chi}_{\dot{\alpha}} = \epsilon_{\dot{\alpha} \dot{\beta}} \bar{\chi}^{\dot{\beta}} 
\eeq
where the Levi-Civita symbol is chosen to be
\beq
\epsilon^{\alpha \beta} = \epsilon^{{\dot{\alpha}}{\dot{\beta}} } = - \epsilon_{\alpha \beta} = - \epsilon_{{\dot{\alpha}}{\dot{\beta}} }   = \begin{pmatrix}
0 & 1 \\
-1 & 0
\end{pmatrix} 
\eeq
Contractions of spinorial quantities are given by
\beq
 \chi \cdot \psi = \chi^{\alpha} \psi_{\alpha} 
 = \psi \cdot \chi \, ,  \qquad
 \bar{\chi} \cdot \bar{\psi} = \bar{\chi}_{\dot{\alpha}} \bar{\psi}^{\dot{\alpha}}
  = \bar{\psi} \cdot \bar{\chi}  
  \eeq
Complex conjugation changes the chirality of spinors. The prescription for the signs is
\beq
(\psi^{\alpha})^{\dagger} =  \bar{\psi}^{\dot{\alpha}} \, , \qquad
(\psi_{\alpha})^{\dagger} =  \bar{\psi}_{\dot{\alpha}} \, , \qquad
(\bar{\chi}^{\dot{\alpha}})^{\dagger} =  \chi^{\alpha} \, , \qquad
(\bar{\chi}_{\dot{\alpha}})^{\dagger} =  \chi_{\alpha} 
\label{complex conjugation 3+1 spinors}
\eeq

We use sigma matrices
\beq
\sigma^{M} = (\mathbf{1}, \sigma^i) \, ,
\, \qquad
\bar{\sigma}^{M} = (\mathbf{1},  -\sigma^i)  
\eeq
where we have defined $(\bar{\sigma}^M)^{\dot{\alpha} \alpha} = \epsilon^{\dot{\alpha} \dot{\beta}}\epsilon^{\alpha \beta} (\sigma^M)_{\beta \dot{\beta}} $.
They satisfy the following set of useful identities 
\begin{align}
\nonumber
& (\sigma^M)_{\alpha \dot{\alpha}} (\bar{\sigma}_M)^{\dot{\beta} \beta} = - 2 \delta_{\alpha}^{\,\,\, \beta} \delta_{\dot{\alpha}}^{\,\,\, \dot{\beta}} \, , \quad
  (\sigma^M)_{\alpha \dot{\alpha}} (\sigma_M)_{ \beta  \dot{\beta}} = - 2 \epsilon_{\alpha \beta} \epsilon_{\dot{\alpha} \dot{\beta}} \, , \quad 
  \mathrm{Tr} (\sigma^M \bar{\sigma}^N) = - 2 \eta^{MN} \, , &  \\
  &   \le \sigma^M \bar{\sigma}^N + \sigma^N \bar{\sigma}^M \ri_{\alpha}^{\,\,\, \beta} = - 2 \eta^{MN} \delta_{\alpha}^{\,\,\, \beta}  \, , \qquad
\le \bar{\sigma}^M \sigma^N + \bar{\sigma}^N \sigma^M \ri_{\,\,\, \dot{\alpha}}^{ \dot{\beta}} = - 2 \eta^{MN} \delta_{\,\,\, \dot{\alpha}}^{ \dot{\beta}}    & 
\end{align}

\vskip 5pt
\subsection*{Spinorial derivatives}
In order to manipulate expressions with spinorial objects it is useful to adopt a notation where spinorial indices are manifest.
For the case of vectors and in particular for partial derivatives this is achieved by defining
\beq \label{conv_derivatives}
\p_{\alpha {\dot{\alpha}}} = (\sigma^M)_{\alpha {\dot{\alpha}}} \p_M \, , \qquad
\p^{\alpha {\dot{\alpha}}} = \epsilon^{\alpha \beta} \epsilon^{\dot{\alpha} \dot{\beta}} \p_{\beta \dot{\beta}} = (\bar{\sigma}^M)^{{\dot{\alpha}} \alpha} \p_M \, , \qquad
\p_M = - \frac12 (\bar{\sigma}_M)^{{\dot{\alpha}} \alpha} \p_{\alpha {\dot{\alpha}}}  
\eeq
which in particular imply
\beq
\square \equiv \p^M \p_M = - \frac12 \p^{\alpha \dot{\alpha}} \p_{\alpha \dot{\alpha}} \, , \qquad
\p^{\alpha \dot{\gamma}} \p_{\dot{\gamma} \beta} = - \delta^{\alpha}_{\,\,\, \beta} \, \square 
\eeq
We assign rules for the coordinates consistently with the requirement $ \p_M x^M = \p_{\alpha {\dot{\alpha}}} x^{\alpha {\dot{\alpha}}} =4 $,  that is
\beq
x^{\alpha {\dot{\alpha}}} = - \frac12 (\bar{\sigma}_M)^{{\dot{\alpha}} \alpha} x^M \, , \qquad
x^M = (\sigma^M)_{\alpha {\dot{\alpha}}} x^{\alpha {\dot{\alpha}}} 
\eeq
It follows that $x^2 \equiv x^M x_M = - 2 x^{\alpha \dot{\alpha}} x_{\alpha \dot{\alpha}}$.

Finally, we define partial spinorial derivatives acting on Grassmann variables as
\beq
\p_{\alpha} \theta^{\beta} = \delta_{\alpha}^{\,\,\, \beta} \, , \qquad
\p^{\beta} \theta_{\alpha} = - \delta_{\alpha}^{\,\,\, \beta} \, , \qquad
\bar{\p}_{\dot{\alpha}} \bar{\theta}^{\dot{\beta}} =  \delta_{\dot{\alpha}}^{\,\,\, {\dot{\beta}}} \, , \qquad
\bar{\p}^{\dot{\beta}} \bar{\theta}_{\dot{\alpha}} = - \delta^{\dot{\alpha}}_{\,\,\, {\dot{\beta}}} 
\eeq
Imposing the reality of $ \delta_M^{\,\, N} = [\p_M , x^N]  $ and $ \delta_{\alpha}^{\,\, \beta} = \lbrace \p_{\alpha} , \theta^{\beta} \rbrace $  we find that spacetime derivatives are anti-hermitian, $ (\p_{M})^{\dagger} = - \p_{M} , $ while the spinorial ones are hermitian, $ (\p_{\alpha})^{\dagger} = \bar{\p}_{\dot{\alpha}}$.  

\subsection*{Superspace}
 
The SUSY generators can be written as
\beq
P_{\alpha {\dot{\alpha}}} = - i \p_{\alpha {\dot{\alpha}}} \, , \qquad
{\cal Q}_{\alpha} = i \le \p_{\alpha} + \frac{i}{2} \bar{\theta}^{\dot{\alpha}}\p_{\alpha {\dot{\alpha}}} \ri \, , \qquad
\bar{\cal Q}_{\dot{\alpha}} =  - i \le \bar{\p}_{\dot{\alpha}} + \frac{i}{2} \theta^{\alpha} \p_{\alpha {\dot{\alpha}}} \ri  
\eeq
such that the algebra is
$\lbrace {\cal Q}_{\alpha}, \bar{\cal Q}_{\dot{\alpha}} \rbrace =
  i \p_{\alpha {\dot{\alpha}}} =  - P_{\alpha {\dot{\alpha}}}$.
The covariant differential operators which anticommute with the supercharges are
\beq
{\cal D}_{\alpha} =  \p_{\alpha} - \frac{i}{2} \bar{\theta}^{\dot{\alpha}}\p_{\alpha {\dot{\alpha}}} =
 -i {\cal Q}_{\alpha} - i \bar{\theta}^{\dot{\alpha}}\p_{\alpha {\dot{\alpha}}}  \, , \qquad
\bar{\cal D}_{\dot{\alpha}} =  \bar{\p}_{\dot{\alpha}} - \frac{i}{2} \theta^{\alpha} \p_{\alpha {\dot{\alpha}}} = 
i \bar{\cal Q}_{\dot{\alpha}} -  i \theta^{\alpha} \p_{\alpha {\dot{\alpha}}}   
\eeq
and they satisfy
\beq
\lbrace {\cal D}_{\alpha}, \bar{\cal D}_{\dot{\alpha}} \rbrace = -  i \p_{\alpha {\dot{\alpha}}} =   P_{\alpha {\dot{\alpha}}}  
\eeq
With these conventions, we obtain
\beq
\bar{\cal Q}_{\dot{\alpha}} = ({\cal Q}_{\alpha})^{\dagger} \, , \qquad
\bar{\cal D}_{\dot{\alpha}} = ({\cal D}_{\alpha})^{\dagger} 
\eeq

We define
\beq
{\cal Q}^2 \equiv \frac12 {\cal Q}^{\alpha} {\cal Q}_{\alpha} \, , \qquad
\bar{\cal Q}^2 \equiv \frac12 \bar{\cal Q}_{\dot{\alpha}} \bar{\cal Q}^{\dot{\alpha}} \, , \qquad
{\cal D}^2 \equiv \frac12 {\cal D}^{\alpha} {\cal D}_{\alpha} \, , \qquad
\bar{\cal D}^2 \equiv \frac12 \bar{\cal D}_{\dot{\alpha}} \bar{\cal D}^{\dot{\alpha}} 
\eeq

\vskip 5pt
\subsection*{Chiral superfields}
Chiral superfields satisfy
$\bar{\cal D}_{\dot{\alpha}} \Sigma (x^M, \theta^{\alpha}, \bar{\theta}^{\dot{\alpha}} ) = 0 $,
and can be written as
\beq
\Sigma (x_L, \theta, \bar{\theta}) = \phi (x_L) + \theta^{\alpha} \psi_{\alpha} (x_L) - \theta^2 F(x_L) \, ,
\qquad
x_L^{\alpha \dot{\alpha}} \equiv x^{\alpha \dot{\alpha}} - i \theta^{\alpha} \bar{\theta}^{\dot{\alpha}}  
\eeq
Similarly, anti-chiral superfields satisfy
${\cal D}_{\alpha} \bar{\Sigma} (x^M, \theta^{\alpha}, \bar{\theta}^{\dot{\alpha}} ) = 0 $,
whose solution is
\beq
\bar{\Sigma} (x_R, \theta, \bar{\theta}) = 
\bar{\phi}(x_R) +  \bar{\theta}_{\dot{\alpha}} \bar{\psi}^{\dot{\alpha}} (x_R) - \bar{\theta}^2 \bar{F}(x_R) \, ,
\qquad
x_R^{\alpha \dot{\alpha}} \equiv x^{\alpha \dot{\alpha}} + i \theta^{\alpha} \bar{\theta}^{\dot{\alpha}} 
\eeq
Using definitions $
\theta^2 \equiv \frac12 \theta^{\alpha} \theta_{\alpha}$, $\bar{\theta}^2 \equiv \frac12 \bar{\theta}_{\dot{\alpha}} \bar{\theta}^{\dot{\alpha}}$, 
we find the following compact expression for the components of a chiral superfield
\beq\label{components}
\phi = \Sigma| \, , \qquad
\psi_{\alpha} = {\cal D}_{\alpha} \Sigma| \, , \qquad
F= {\cal D}^2 \Sigma|  
\eeq
where $ | $ means that we evaluate the expression at $ \theta= \bar{\theta} = 0$. The anti-chiral components are simply given by the complex conjugated of 
these expressions.

\subsection*{Pauli matrices in light-cone coordinates}
Pauli matrices matrices in light-cone coordinates are
\bea
\sigma^{\pm} = \frac{1}{\sqrt{2}} (\sigma^{3} \pm \sigma^{0}) \, , &&\qquad
\bar{\sigma}^{\pm} = \frac{1}{\sqrt{2}} (\bar{\sigma}^{3} \pm \bar{\sigma}^{0}) 
\nl
\sigma^- = - \bar{\sigma}^+ = \sqrt{2}
\begin{pmatrix}
0 & 0 \\
0 & -1 
\end{pmatrix} \, , &&\qquad
\sigma^+ = - \bar{\sigma}^-= \sqrt{2}
\begin{pmatrix}
1 & 0 \\
0 & 0 
\end{pmatrix} 
\eea
Therefore, for instance we write (from \eqref{conv_derivatives})
\beq
\begin{aligned} \label{light-cone derivatives}
& \partial_{\alpha \dot{\alpha}} = (\sigma^+)_{\alpha \dot{\alpha}} \, \partial_+ +  (\sigma^-)_{\alpha \dot{\alpha}} \, \partial_- +  (\sigma^1)_{\alpha \dot{\alpha}} \, \partial_1 +   (\sigma^2)_{\alpha \dot{\alpha}} \, \partial_2 \\
& \partial^{\alpha \dot{\alpha}} = (\bar{\sigma}^+)^{\dot{\alpha}\alpha } \, \partial_+ +  (\bar{\sigma}^-)^{\dot{\alpha}\alpha } \, \partial_- +  (\bar{\sigma}^1)^{\dot{\alpha}\alpha }\, \partial_1 +   (\bar{\sigma}^2)^{\dot{\alpha}\alpha } \, \partial_2 
\end{aligned}
\eeq
with $\partial_{\pm} = \frac{1}{\sqrt{2}} (\partial_3 \pm \partial_0)$.
\\

\subsection*{Conventions in 2+1 dimensions}

Non-relativistic quantities in 2+1 dimensions are obtained from the null reduction of 3+1 dimensional Minkowski spacetime.
The prescription is to introduce light-cone coordinates $ x^M = (x^-, x^+, x^1, x^2) = (x^-, x^{\mu}) , $ compactify along a small circle in the $ x^- $ direction and perform the identifications
\beq
\p_- \rightarrow im \, , \qquad 
\p_+ \rightarrow \p_t \, , \qquad
 \phi (x^M) = e^{i m x^-} \varphi (x^{\mu}) 
\eeq
where $ m $ is the adimensional eigenvalue of the $ U(1) $ mass operator and $ \phi(x^M) $ is a local field.

Non-relativistic fermions in 2+1 dimensions are parametrized by two complex Grassmann scalars $\xi(x^{\mu})$ and $\chi(x^{\mu})$.
Under null reduction the identification with the 3+1 dimensional left-handed Weyl spinor $ \psi(x^M) $ works as follows
\beq
\psi_{\alpha} (x^M) = e^{i m x^-} \tilde{\psi}_{\alpha} (x^{\mu}) =e^{i m x^-} 
\begin{pmatrix}
\xi (x^{\mu}) \\
\chi (x^{\mu})
\end{pmatrix}
\eeq
Under complex conjugation we choose the prescription
\beq
\bar{\psi}_{\dot{\alpha}} =
 (\psi_{\alpha})^{\dagger} = 
 e^{- i m x^-} (\tilde{\psi}_{\alpha})^{\dagger} \equiv e^{- i m x^-}  \begin{pmatrix}
\bar{\xi} (x^{\mu}) \\
 \bar{\chi} (x^{\mu})
\end{pmatrix}
\eeq
Using identities \eqref{raise&lower} it is easy to infer the identification with the components of $\psi^\alpha$ and $\bar{\psi}^{\dot{\alpha}}$.

Taking the mass as a dimensionless parameter enforces the energy dimensions of the non-relativistic fermion to be
\beq
[\xi] = E^2 \, , \qquad
[\chi] = E 
\eeq
These assignments immediately follow when performing the null reduction of the Weyl Lagrangian
\beq
\mathcal{L} = i \psi^{\dagger} \bar{\sigma}^M \p_M \psi \rightarrow
\sqrt{2} m \bar{\xi} \xi + \sqrt{2} i \bar{\chi} \p_t \chi - i \bar{\xi} (\p_1 - i \p_2) \chi - i \bar{\chi} (\p_1 + i \p_2) \xi 
\eeq
We observe that the only dynamical component is $ \chi $, while $\xi$ turns out to be an auxiliary field that can be integrated out from the action.

Since null reduction affects only space-time coordinates without modifying the spinorial ones, we obtain ${\cal N}=2$ supersymmetry in three dimensions. According to the ordinary pattern for which the three dimensional ${\cal N}=2$ superspace is ``equal'' to the four-dimensional ${\cal N}=1$ superspace, all the properties related to manipulations of covariant derivatives and supercharges, \emph{e.g.} the D-algebra procedure, are directly inherited from the (3+1) relativistic theory under a suitable re-interpretation of the spinorial objects. 

In particular, the algebra of covariant derivatives reads
\beq
\{ D_\alpha , \bar{D}_\beta \} = - i \p_{\alpha \beta}\, ,  \qquad \qquad \{ D^\alpha , \bar{D}^\beta \} = - i \p^{\beta \alpha}
\eeq
where, as follows from \eqref{light-cone derivatives}, the three-dimensional derivatives are given by 
\beq\label{3d_derivatives}
\partial_{\alpha \beta} = \begin{pmatrix}
\sqrt{2} \p_t  &   \p_1 - i \p_2 \\
 \p_1 + i \p_2 & - i \sqrt{2} M
\end{pmatrix} 
\qquad \quad 
\partial^{\alpha \beta} = \begin{pmatrix}
- i\sqrt{2} M  &   -(\p_1 - i \p_2) \\
 -(\p_1 + i \p_2) & \sqrt{2} \p_t 
\end{pmatrix} 
\eeq
They satisfy the following identities
\beq 
\p^{\alpha \beta} = \epsilon^{\alpha \delta} \epsilon^{\beta \gamma} \p_{\gamma \delta} 
\qquad \qquad
\p_{\beta \alpha} = \epsilon_{\alpha \gamma} \epsilon_{\beta \delta} \, \p^{\gamma \delta}
\eeq
Therefore, we have for instance $\bar{\xi}_{\alpha} \p^{\alpha \beta} \chi_\beta = \bar{\xi}^{\alpha} \p_{\beta \alpha} \chi^\beta$.
Identities which turn out to be useful for the reduction of the action to components are
\beq \label{useful_identities}
\begin{aligned}
& [ D^\alpha , \bar{D}^2 ] = i \p^{\beta\alpha } \bar{D}_\beta \, , \qquad \quad [ \bar{D}^\alpha , D^2 ] = -i \p^{\alpha \beta} D_\beta \\
& D^2 \bar{D}^2 + \bar{D}^2 D^2 = (2iM\p_t + \p_i^2) + D^\alpha \bar{D}^2 D_\alpha = (2iM\p_t + \p_i^2)  + \bar{D}_\alpha D^2 \bar{D}^\alpha
\end{aligned}
\eeq

\section{Non-relativistic Wess-Zumino model in components} \label {sec_components}

In this appendix we perform the reduction to components of the non-relativistic superaction \eqref{non-rel WZ action in superfield formalism} and show how to obtain expression \eqref{Lagrangiana finale WZ interagente potenziale cubico} after eliminating the auxiliary fields. 

For simplicity we focus only on the Kahler term. Applying the prescription for the Berezin integration \eqref{Berezin integration null reduction} we can write 
\beq
S_{kin} = \int d^3 x \, \left[\bar{D}^2 \bar{\Sigma}_a \, D^2 \Sigma_a + \bar{D}_{\alpha} \bar{\Sigma}_a \; \bar{D}^{\alpha} D^2 \Sigma_a + \bar{\Sigma}_a \, \bar{D}^2 D^2 \Sigma_a  \right] 
\eeq
where the non-relativistic covariant derivatives are given in eq. \eqref{nonrelQD} and $  a=1,2 $ labels the two sectors of superfields.

We define the theta expansion of the Wess-Zumino chiral superfields as (here $\theta^1, \theta^2$ indicate components 1 and 2 of the $\theta^\alpha$ spinor)
\beq \label{sigma_components}
\begin{aligned}
& \Sigma_1 = \varphi_1 + \theta^1 \xi_1 + \theta^2 \, 2^{\frac14}\sqrt{m} \, \chi_1 - \frac12 \theta^\alpha \theta_\alpha F_1 \\
& \Sigma_2 = \varphi_2 + \theta^1 \xi_2 + \theta^2 \, i 2^{\frac14} \sqrt{2m}  \, \chi_2 - \frac12 \theta^\alpha \theta_\alpha F_2 
\end{aligned}
\eeq
where a convenient rescaling of the grassmannian fields has been implemented in order to have the standard normalization of the kinetic terms, with $\varphi_a$ and $\chi_a$ sharing the same dimensions. Using these conventions we obtain
\beq
\begin{aligned}
S_{kin}  = & \int d^3 x \,  \left[ 2 i m \bar{\varphi}_1 (\p_t \varphi_1) + \bar{\varphi}_1 \p_i^2 \varphi_1 - 4 i m \bar{\varphi}_2 (\p_t \varphi_2) + \bar{\varphi}_2 \p_i^2 \varphi_2 + \bar{F}_1 F_1 + \bar{F}_2 F_2
\right. \\
& \left. + \sqrt{2} m \bar{\xi}_1 \xi_1 + 2 i m \bar{\chi}_1 (\p_t \chi_1) -  2^{\frac{1}{4}} i \sqrt{m} \, \bar{\xi}_1 (\p_1 - i \p_2) \chi_1  
  -  2^{\frac{1}{4}} i \sqrt{m} \bar{\chi}_1 (\p_1 + i \p_2) \xi_1 \right. 
\\ & \left. - 2 \sqrt{2} m \bar{\xi}_2 \xi_2 + 4 i m \bar{\chi}_2 (\p_t \chi_2) +  2^{\frac{1}{4}}  \sqrt{2m} \, \bar{\xi}_2 (\p_1 - i \p_2) \chi_2  
  -  2^{\frac{1}{4}}  \sqrt{2m} \bar{\chi}_2 (\p_1 + i \p_2) \xi_2 \right]
\end{aligned}
\label{Azione modello WZ interagente al primo ordine}
\eeq
Integrating out the non-dynamical fields $ F_{1,2} $ and $ \xi_{1,2} $ we find
\beq
\begin{aligned}
S_{kin}  =  \int d^3 x & \, \left[   2 i m \bar{\varphi}_1 \p_t \varphi_1 + \bar{\varphi}_1 \p_i^2 \varphi_1 
- 4 i m \bar{\varphi}_2 \p_t \varphi_2   + \bar{\varphi}_2 \p_i^2 \varphi_2   \right. \\
& \; \left. + 2 i m  \bar{\chi}_1 \p_t \chi_1 +  \bar{\chi}_1 \p_i^2 \chi_1 + 4 i m \bar{\chi}_2 \p_t \chi_2
 -  \bar{\chi}_2 \p_i^2 \chi_2 \right]
\end{aligned}
\eeq  
The component form of the interacting part of the action can be similarly obtained by means of the standard superspace manipulations combined with the Berezin integration \eqref{Berezin integration null reduction}.
The result in terms of dynamical fields is
\beq
\begin{aligned}
S_{int}   = &  \int d^3x  \;  \Big[- 4 |g|^2 |\varphi_1\varphi_2|^2 -  |g|^2 |\varphi_1|^4 \\
&  -  i g \left( \sqrt{2} \varphi_1 \chi_1 (\p_1 - i \p_2) \bar{\chi}_2 -2 \bar{\varphi}_2 \chi_1 (\p_1 - i \p_2) \chi_1 + 2 \sqrt{2} \varphi_1 ((\p_1 - i \p_2)\chi_1) \bar{\chi}_2  \right) + {\rm h.c.} \\
& + 2 |g|^2 \left( -|\varphi_1|^2 \bar{\chi}_1 \chi_1 - 4 |\varphi_1|^2 \bar{\chi}_2 \chi_2 + 2 |\varphi_2|^2 \bar{\chi}_1 \chi_1 + 2 \sqrt{2} \varphi_1 \varphi_2 \bar{\chi}_1 \bar{\chi}_2 + 2 \sqrt{2} \bar{\varphi}_1 \bar{\varphi}_2 \chi_2 \chi_1 \right) 
     \Big]  \\ &
\end{aligned}   
\label{Lagrangiana finale WZ interagente potenziale cubico}
\eeq


\section{Quantum corrections in components}
\label{app-nonrel WZ model in components}

As a cross-check of the results obtained in section \ref{sec-Non-relativistic Wess-Zumino model} using superspace
formalism, in this appendix we compute one and two-loop quantum corrections to the self-energies and one-loop corrections to the vertices using the component formalism. 

\begin{figure}[h]
\begin{subfigure}[b]{0.5\linewidth}
\centering
\includegraphics[scale=1]{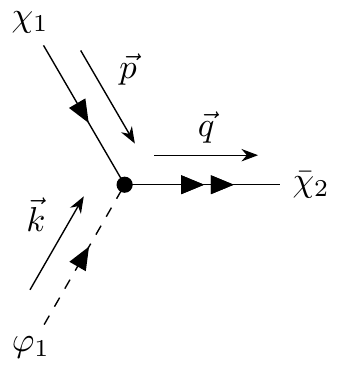}
\subcaption{$ -\sqrt{2} i g \left[ (q_1 - i q_2) -2 (p_1 - i p_2) \right] $}
\end{subfigure}
\begin{subfigure}[b]{0.5\linewidth}
\centering
\includegraphics[scale=1]{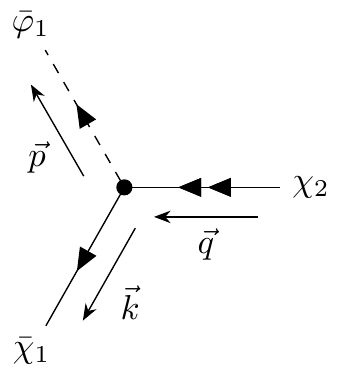}
\subcaption{$ \sqrt{2} i g^{*} \left[ (q_1 + i q_2) -2 (k_1 + i k_2) \right] $}
\end{subfigure}
\begin{subfigure}[b]{0.5\linewidth}
\centering
\includegraphics[scale=1]{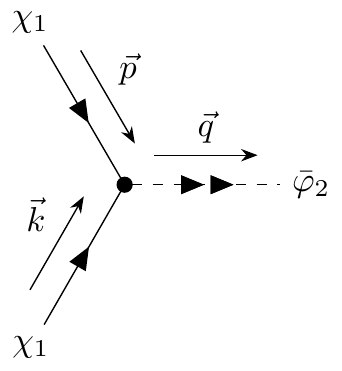}
\subcaption{$ -2 i g \left[ (k_1 - i k_2) - (p_1 - i p_2) \right] $}
\end{subfigure}
\begin{subfigure}[b]{0.5\linewidth}
\centering
\includegraphics[scale=1]{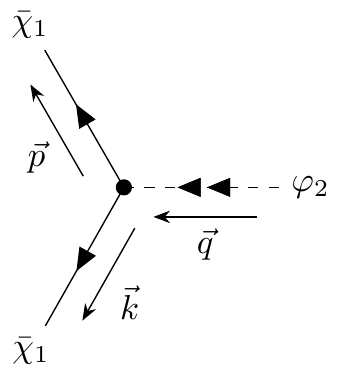}
\subcaption{$ 2 i g^{*} \left[ (k_1 + i k_2) - (p_1 + i p_2) \right] $}
\end{subfigure}
\caption{Feynman rules for three-point vertices. Scalars are denoted by dashed lines, while fermions by continuous lines. }
\label{fig18_Vertici_3_Lagrangiana_campi_dinamici_componenti}
\end{figure}

\begin{figure}[h]
\centering
\begin{subfigure}[b]{0.3\linewidth}
\includegraphics[scale=1]{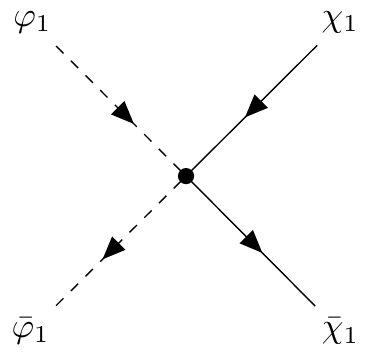}
\subcaption{$ -2 i |g|^2 $}
\end{subfigure}
\begin{subfigure}[b]{0.3\linewidth}
\includegraphics[scale=1]{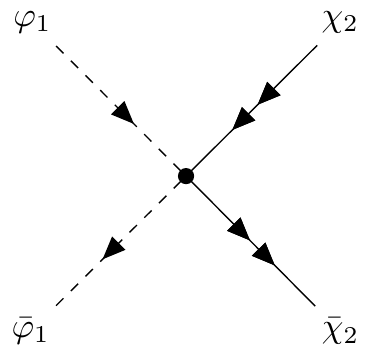}
\subcaption{$ -8 i |g|^2 $}
\end{subfigure}
\begin{subfigure}[b]{0.3\linewidth}
\includegraphics[scale=1]{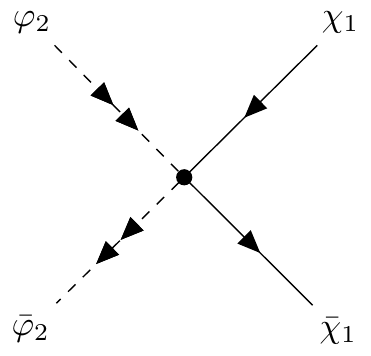}
\subcaption{$ 4 i |g|^2 $}
\end{subfigure}
\begin{subfigure}[b]{0.3\linewidth}
\includegraphics[scale=1]{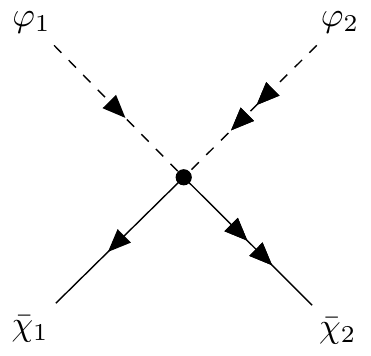}
\subcaption{$ 4 \sqrt{2} i |g|^2 $}
\end{subfigure}
\begin{subfigure}[b]{0.3\linewidth}
\includegraphics[scale=1]{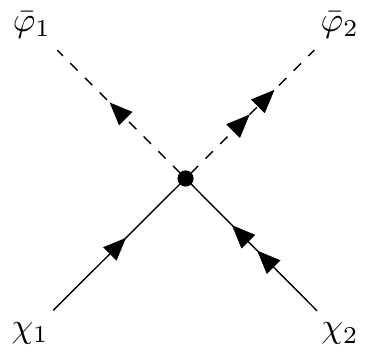}
\subcaption{$ 4 \sqrt{2} i |g|^2 $}
\end{subfigure}
\begin{subfigure}[b]{0.3\linewidth}
\includegraphics[scale=1]{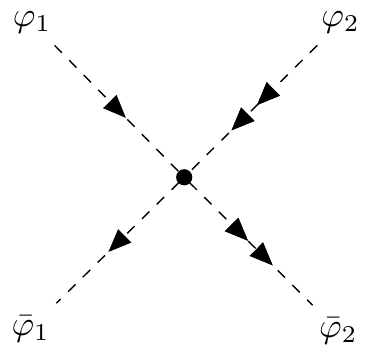}
\subcaption{$ -4  i |g|^2 $}
\end{subfigure}
\begin{subfigure}[b]{0.3\linewidth}
\includegraphics[scale=1]{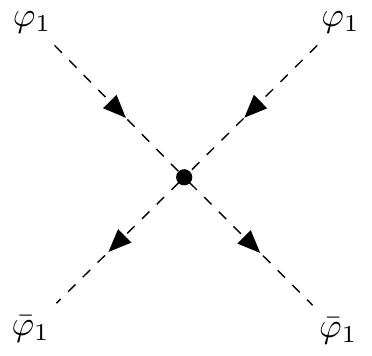}
\subcaption{$ -4 i |g|^2 $}
\end{subfigure}
\caption{Feynman rules for four-point vertices. Scalars are denoted by dashed lines, while fermions by continuous lines.}
\label{fig19_Vertici_4_Lagrangiana_campi_dinamici_componenti}
\end{figure}

We start from the action in components \eqref{Lagrangiana finale WZ interagente potenziale cubico}. Scalars and fermions share the same kinetic operator. Therefore, the tree-level propagators are
\bea
&& \langle \varphi_{1}(\omega, \vec{p}) \bar{\varphi}_{1}(-\omega, -\vec{p}) \rangle = \langle \chi_{1}(\omega, \vec{p}) \bar{\chi}_{1}(-\omega, -\vec{p}) \rangle = 
\frac{i}{2m \omega - \vec{p}^{\, 2} + i \varepsilon}  \nonumber \\
&& \langle \varphi_{2}(\omega, \vec{p}) \bar{\varphi}_{2}(-\omega, -\vec{p}) \rangle = \langle \chi_{2}(\omega, \vec{p}) \bar{\chi}_{2}(-\omega, -\vec{p}) \rangle = 
\frac{i}{4m \omega - \vec{p}^{\, 2} + i \varepsilon}  
\eea
Interaction vertices can be read directly from the lagrangian and are shown in fig. \ref{fig18_Vertici_3_Lagrangiana_campi_dinamici_componenti} and \ref{fig19_Vertici_4_Lagrangiana_campi_dinamici_componenti}, where we use dashed and continous lines to denote scalars and fermions, respectively.
 
In order to classify the admitted diagrams, we can take into account that the reduction in components does not affect the propagators as functions of  $\omega$ and $\vec{p}$. Therefore, the arguments that led to formulate the fundamental selection rule \ref{srule2} are still true. Moreover, the conservation of particle number at each vertex still provides the driving rule to select the admissible topologies and arrows configurations.
 
In order to properly define physical quantities and Green functions, we introduce renormalized fields and couplings defined as
\beq
\begin{cases}\label{ren_components}
\varphi_a = Z_a^{-1/2} \,  \varphi_a^{(B)} = \le 1- \frac12 \delta_{\varphi_a} \ri \varphi_a^{(B)} \qquad a=1,2\\
\chi_a = Z_a^{-1/2} \,  \chi_a^{(B)} = \le 1- \frac12 \delta_{\chi_a} \ri \chi_a^{(B)}  \\
m = Z^{-1}_m m^{(B)} = (1 - \delta_m) m^{(B)} \\
g = \mu^{-\varepsilon} Z^{-1}_g g^{(B)} = \mu^{-\varepsilon} (1 - \delta_g) g^{(B)} 
\end{cases}
\eeq
Spatial integrals are computed in dimension $ d = 2 - \varepsilon $ and we have introduced the mass scale $ \mu $ to keep the coupling constant dimensionless.

\subsubsection*{One-loop corrections to the self-energies}

\begin{figure}[h]
\centering
\begin{subfigure}[b]{0.45\linewidth}
\centering
\includegraphics[scale=1.35]{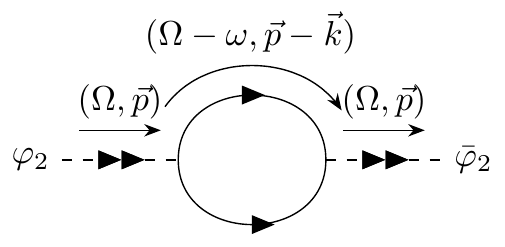}
\subcaption{$ i \mathcal{M}^{(2)}_{\mathrm{b}} (\varphi_2, \bar{\varphi}_2) $}
\end{subfigure}
\begin{subfigure}[b]{0.45\linewidth}
\centering
\includegraphics[scale=1.35]{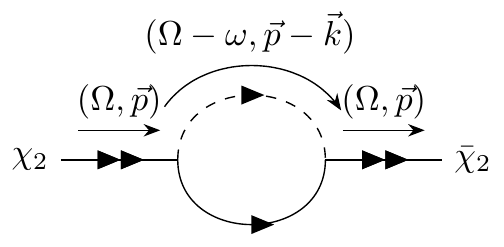}
\subcaption{$ i \mathcal{M}^{(2)}_{\mathrm{f}} (\chi_2, \bar{\chi}_2) $}
\end{subfigure}
\caption{1-loop correction to the scalar (a) and fermionic (b) self-energies in sector 2.}
\label{fig20_Correzione 1-loop settore 2 dinamico}
\end{figure}

By applying selection rule \ref{srule2} and particle number conservation there are no admissible one-loop self-energy diagrams for particles in sector 1.   
For fields in sector 2, instead we find a non-vanishing contribution both for the scalar and the fermion corresponding to the diagrams in fig. \ref{fig20_Correzione 1-loop settore 2 dinamico}.
Direct inspection leads to 
\beq
i \mathcal{M}_{\mathrm{b}}^{(2)}=
i \mathcal{M}_{\mathrm{f}}^{(2)} = \frac{2 |g|^2}{(2 \pi)^3} \int d\omega \, d^2 k \,
\frac{(\vec{p} -2 \vec{k})^2}{\left[2 m \omega - \vec{k}^2 + i \varepsilon \right] \left[2m (\Omega - \omega) - (\vec{p} - \vec{k})^2 + i \varepsilon \right]}  
\eeq
 
We use the residue method to perform the integration in $\omega$, obtaining 
\beq
\mathcal{M}^{(2)} =- \frac{ |g|^2}{m} \int \frac{d^2 k}{(2 \pi)^2} \, 
\frac{(\vec{p} - 2 \vec{k})^2}{2m \Omega - \vec{k}^2 - (\vec{p}-\vec{k})^2 + i \varepsilon} 
\eeq
The remaining integral is UV divergent and can be computed with standard techniques of dimensional regularization. We obtain
\beq
\mathcal{M}^{(2)} =  \frac{|g|^2}{m} \, d \, \frac{ \mu^{2(2-d)}}{(4 \pi)^{d/2}} \, \Gamma \le 
- \frac{d}{2} \ri \, \le \frac{\vec{p}^2}{4} - m \Omega \ri^{\frac{d}{2}} =  \frac{|g|^2}{2 \pi m} \le 2 m \Omega - \frac{\vec{p}^2}{2}  \ri \frac{1}{\varepsilon} + \mathrm{finite}
\label{bubbola}
\eeq
In minimal subtraction scheme the $1/\epsilon$ pole is cancelled by setting in  \eqref{ren_components}
\beq
\delta_{\varphi_2}^{\rm (1loop)}= \delta_{\chi_2}^{\rm (1loop)} = - \frac{|g|^2}{4 \pi m} \frac{1}{\varepsilon} \, , \qquad \; 
\delta_m^{\rm (1loop)} = 0  
\label{controtermini phi2}
\eeq
whereas $\delta_{\varphi_1}^{\rm (1loop)} = \delta_{\chi_1}^{\rm (1loop)} = 0$. This result is consistent with the one-loop renormalization \eqref{superfield_ren} that we have found in superspace.

\subsubsection*{One-loop corrections to three-point vertices}

In the action in components there are two types of three-point vertices (see fig. \ref{fig19_Vertici_4_Lagrangiana_campi_dinamici_componenti}). We  discuss them separately.

Vertex $ \mathbf{V}_3 ( \chi_1 , \chi_1 , \bar{\varphi}_2 )  $ and its complex conjugate are not corrected at one loop since we cannot build any diagram consistent with particle number conservation. It then follows that
\beq
\le \delta_g + \delta_{\chi_1} + \frac12 \delta_{\varphi_2}^* \ri \Big|_{\rm (1loop)}   = 0 
\eeq
Using result (\ref{controtermini phi2}), we then find
\beq
\delta_g^{\rm (1loop)}  =   \frac{|g|^2}{8 \pi m} \frac{1}{\varepsilon}  
\label{controtermini primo vertice a 3}
\eeq

\begin{figure}[h]
\centering
\includegraphics[scale=1.5]{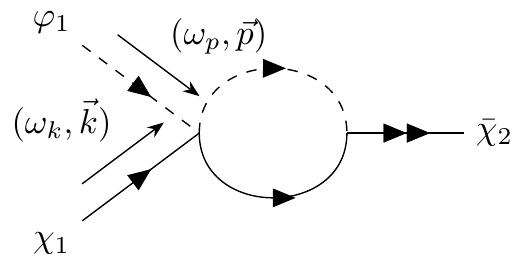}
\caption{1-loop correction to the 3-point vertex.}
\label{fig21_Correzione 1-loop vert a 3}
\end{figure}

Vertex  $ \mathbf{V}_3 ( \varphi_1 , \chi_1 , \bar{\chi}_2 ) $ has in principle a one-loop contribution shown in fig.  \ref{fig21_Correzione 1-loop vert a 3}.

After $\omega$-integration by residues, this diagram gives
\beq
\mathcal{M}^{(3)} (\varphi_1, \chi_1, \bar{\chi}_2) = - \frac{|g|^2}{m}  \frac{ \sqrt{2} g}{(2 \pi)^2}
\int d^2 l \, \frac{(p_1 + k_1) - i (p_2 + k_2) - 2 (l_1 - i l_2)}
{2 m (\omega_p + \omega_k) - \vec{l}^2 - (\vec{p} + \vec{k} - \vec{l})^2 + i \varepsilon}  
\eeq
We perform dimensional regularization along the spatial directions. 
Since the integrand contains in the numerator an expression which  depends explicitly on the $l_i$ components in a non-covariant way, before  continuing the integrand to $d$ dimensions we need to give a prescription to covariantize the numerator. 
We introduce the vector $ \vec{v}=(1,-i)$ and write the numerator as $\vec{v} \cdot (\vec{p} + \vec{k} - 2  \vec{l}\, )$. We then continue the integrand to $d$ dimensions, promoting also $\vec{v}$ to a $d$-dimensional vector with only the first two components different from zero. We have
\beq
\mathcal{M}^{(3)} (\varphi_1, \chi_1, \bar{\chi}_2) =- \frac{|g|^2}{m}  \frac{ \sqrt{2} g \mu^{3(2-d)}}{(2 \pi)^d}
\int d^d l \, \frac{ \vec{v} \cdot (\vec{p} + \vec{k} - 2  \vec{l} \, ) }
{2 m (\omega_p + \omega_k) - \vec{l}^2 - (\vec{p} + \vec{k} - \vec{l})^2 + i \varepsilon}  
\eeq
With the change of variables $\vec{q} = \vec{l} - \frac{\vec{p}+ \vec{k}}{2}$ we obtain
\beq
\mathcal{M}^{(3)} (\varphi_1, \chi_1, \bar{\chi}_2) = - \frac{|g|^2}{m} \frac{ \sqrt{2} g  \mu^{3(2-d)}}{(2 \pi)^d}
\int d^d q \, \frac{\vec{v} \cdot \vec{q}}{ q^2 - m (\omega_p + \omega_k) + \frac{(\vec{p} + \vec{k})^2 }{4} + i \varepsilon}  
\eeq
This integral vanishes for symmetry reasons. The lack of one-loop corrections then implies that conterterms in eq. \eqref{ren_components} need to satisfy
\beq
\le \delta_g + \frac12 \delta_{\varphi_1} + \frac12 \delta_{\chi_1} + \frac12 \delta_{\chi_2} \ri \Big|_{\rm (1loop)}= 0 
\eeq
This condition is automatically satisfied by results in eqs. (\ref{controtermini phi2}, \ref{controtermini primo vertice a 3}).

We note that the one-loop result $\delta_g = - \frac12 \delta_{\chi_2}$ is the component version of the superspace constraint \eqref{ren_constraint}. As expected, quantum corrections do not break supersymmetry.

\subsubsection*{One-loop corrections to four-point vertices}

In principle, the one-loop evaluation of self-energies and three-point vertices allows to solve for all the unknowns in lagrangian 
(\ref{Lagrangiana finale WZ interagente potenziale cubico}). Moreover, we have verified that the corrections are all consistent between themselves and with the superspace results. However, we may want to consider  some 1PI diagrams involving four-point vertices, in order to provide further evidence that SUSY is preserved also working in components.

\begin{figure}[h]
\centering
\begin{subfigure}[b]{0.3\linewidth}
\includegraphics[scale=1]{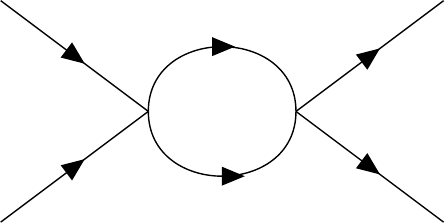}
\end{subfigure}
\begin{subfigure}[b]{0.3\linewidth}
\includegraphics[scale=1]{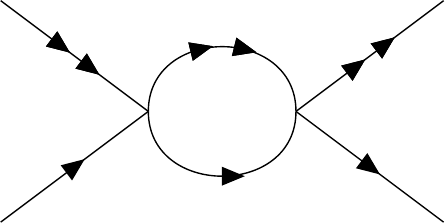}
\end{subfigure}
\begin{subfigure}[b]{0.3\linewidth}
\includegraphics[scale=1]{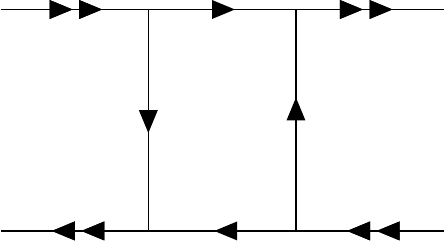}
\end{subfigure}
\begin{subfigure}[b]{0.3\linewidth}
\includegraphics[scale=1]{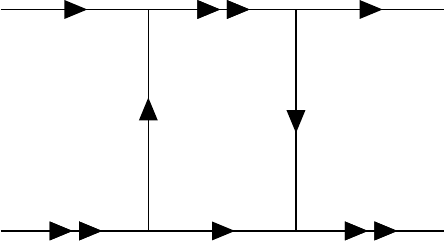}
\end{subfigure}
\begin{subfigure}[b]{0.3\linewidth}
\includegraphics[scale=1]{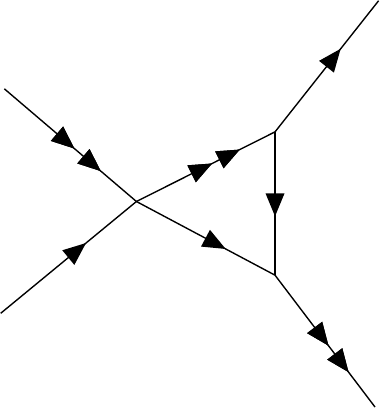}
\end{subfigure}
\begin{subfigure}[b]{0.3\linewidth}
\includegraphics[scale=1]{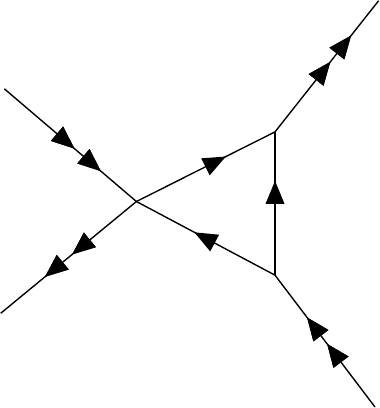}
\end{subfigure}
\begin{subfigure}[b]{0.3\linewidth}
\includegraphics[scale=1]{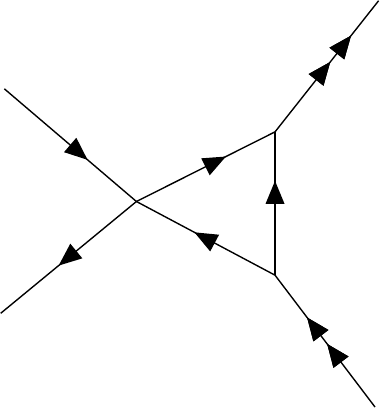}
\end{subfigure}
\caption{Possible topologies of one-loop corrections to four-point vertices for the dynamical fields.
In the picture we do not distinguish between bosonic and fermionic lines.}
\label{fig22_Classificazione_topologie_diagrammi_4}
\end{figure}

Compared to the previous cases, we have far more possibilities to build four-point diagrams with the vertices at our disposal (see figs. \ref{fig18_Vertici_3_Lagrangiana_campi_dinamici_componenti}, \ref{fig19_Vertici_4_Lagrangiana_campi_dinamici_componenti}). All the topologies of diagrams consistent with particle number conservation at each vertex are reported in fig. \ref{fig22_Classificazione_topologie_diagrammi_4}.

\begin{figure}[h]
\centering
\begin{subfigure}[b]{0.5\linewidth}
\centering
\includegraphics[scale=1.3]{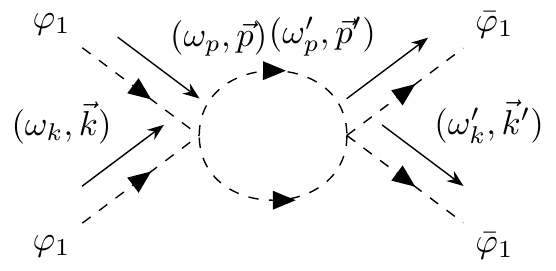}\caption{}
\end{subfigure}
\caption{One-loop 1PI correction to the four-point vertex with external scalars from sector 1, coming from the \emph{s} channel.}
\label{fig23_Correzioni canali t e u 1-loop vert a 4}
\end{figure}

For example, we consider the first diagram in fig. \ref{fig22_Classificazione_topologie_diagrammi_4}, \emph{i.e.} the one-loop correction to the vertex $ \mathbf{V}_4 (\varphi_1 , \varphi_1 , \bar{\varphi}_1 , \bar{\varphi}_1)$. This is the only diagram among the many containing as external lines only fields from sector 1. 
We report the precise assignments of momenta and energy in fig. \ref{fig23_Correzioni canali t e u 1-loop vert a 4}.

The \emph{t} and \emph{u}-channel diagrams vanish because we have circulating arrows in the internal loop.
After integration in $\omega$, the integral corresponding to the \emph{s}-channel diagram is
\beq
\mathcal{M}^{(4)} (\varphi_1, \varphi_1, \bar{\varphi}_1, \bar{\varphi}_1) = - \frac{4 |g|^4}{m} \int \frac{d^2 l}{(2 \pi)^2} \,
\frac{1}{2m (\omega_p + \omega_k) - \vec{l}^2 - (\vec{p} + \vec{k} - \vec{l})^2 + i \varepsilon} 
\eeq
Performing the change of variables $\vec{q} = \vec{l} - \frac{\vec{p} + \vec{k}}{2}$, in dimensional regularization we can write
\beq
\mathcal{M}^{(4)} (\varphi_1, \varphi_1, \bar{\varphi}_1, \bar{\varphi}_1) =  \frac{4|g|^4}{m} \frac{ \mu^{4(2-d)}}{(4 \pi)^{d/2}} \, \frac{1}{\Gamma(d/2)} \int_0^{\infty} dq \,
\frac{q^{d-1}}{q^2 -m (\omega_p + \omega_k) + \frac{(\vec{p}+ \vec{k})^2}{4} + i \varepsilon} 
\eeq
After performing the last integration and expanding in $ \varepsilon=2-d $ we find
\beq
\mathcal{M}^{(4)} (\varphi_1, \varphi_1, \bar{\varphi}_1, \bar{\varphi}_1) =  \frac{|g|^4}{\pi m} \frac{1}{\varepsilon} + \mathrm{finite} 
\label{correzione 1 loop vertice a 4 scalari}
\eeq
The renormalization condition in minimal subtraction scheme requires  
\beq
\mathcal{M}^{(4)} (\varphi_1, \varphi_1, \bar{\varphi}_1 \bar{\varphi}_1) - 4  |g|^2 (2\delta_{g} + 2 \delta_{\varphi_1}) = 0
\eeq
We then obtain
\beq
\delta_g^{\rm (1loop)} =  \frac{|g|^2}{8 \pi m} \frac{1}{\varepsilon} 
\eeq
which is consistent with \eqref{controtermini primo vertice a 3}.

This confirms that SUSY is preserved by quantum corrections. Moreover,
the quantum corrections of the coupling constant  $g$ 
are completely determined by the wave-function renormalization, as
expected from the non-renormalization theorem.

\subsubsection*{Two-loop corrections to the self-energy}

In component field formalism the number of Feynman diagrams at each loop order is much greater than using the superspace approach.
This makes the evaluation of quantum corrections more involved when the number of loops increases.
However, in the non-relativistic case selection rules \ref{srule2} and \ref{srule3} help in drastically decreasing the number of diagrams to be considered.
In particular self-energies are easily treatable also at two loops. Here we report this calculation as an example of higher loop corrections in component field formalism.

\begin{figure}[h]
\centering
\includegraphics[scale=1.5]{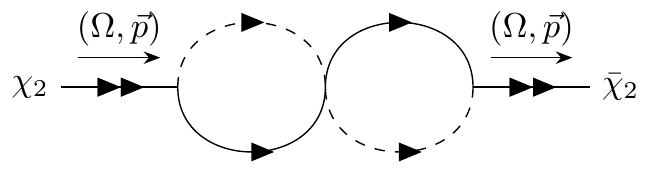}
\caption{Two-loop correction to the self-energy for the dynamical fermion in sector 2.}
\label{fig24_correzione 2 loop fermione dinamico}
\end{figure}

At two loops the only self-energy diagram compatible with the selections rules is the one for the $\chi_2$ fermion, depicted in fig. \ref{fig24_correzione 2 loop fermione dinamico}. Since there is no possibility to draw a non-vanishing diagram for the corresponding scalar, consistency with SUSY invariance requires this contribution to vanish. We now prove that this is indeed the case.

Writing down the corresponding integral and first performing the $ \omega_k, \omega_l $ integrations by using the residue technique we find
\beq
\mathcal{M}^{(4)}_{\mathrm{f}} (\chi_2, \bar{\chi}_2) =
- \frac{|g|^4}{m^2} \int \frac{d^2 k \, d^2 l}{(2 \pi)^4} \, 
\frac{\vec{p}^{\, 2} + 4 \vec{l} \cdot \vec{k} + 2 (\vec{l}+ \vec{k}) \cdot \vec{p}}
{\left[2 m \Omega - \vec{k}^2 - (\vec{p}- \vec{k})^2 + i \varepsilon \right]\left[2 m \Omega - \vec{l}^2 - (\vec{p}- \vec{l})^2 + i \varepsilon \right]}  
\eeq
Performing the change of variables 
\beq
\vec{k}= \vec{K}+ \frac{\vec{p}}{2} \, , \qquad
\vec{l}= \vec{L}+ \frac{\vec{p}}{2}  
\eeq
and continuing the integral to $d=2-\epsilon$ dimensions we find
\beq
\mathcal{M}^{(4)}_{\mathrm{f}} (\chi_2, \bar{\chi}_2) =
- \frac{|g|^4}{m^2} \frac{\mu^{4(2-d)}}{(2 \pi)^{2d}} \int d^d K \, d^d L \, \frac{4 \vec{K} \cdot \vec{L}}
{\left[2 m \Omega - 2 K^2 - \frac{\vec{p}^2}{4} + i \varepsilon \right]\left[2 m \Omega - 2 L^2 - \frac{\vec{p}^2}{2} + i \varepsilon \right]} 
\eeq
The two integrals vanish for symmetry reasons.


\section{Example of non-relativistic supergraph calculation} \label{examples}

As an example of supergraph calculation we consider the two-loop corrrection to the three-point vertex.
 
Conservation of the particle number at each vertex restricts the allowed diagrams to the single non-planar graph of   fig. \ref{fig14_2 loop supervertici a 3}, where we have depicted all possible consistent assignments of arrows.
 
This is a case in which the number of chiral and anti-chiral vertices is different. Consequently, the factors of  covariant derivatives are not only used to simplify propagators, but as the result of applying D-algebra \eqref{rules covariant derivatives giving momenta in supergraphs 2}, they give powers of momenta at the numerator which might affect the convergence of the $\omega$ integrations. Therefore, though all the diagrams contain closed loops of arrows and they should vanish due to selection rule \ref{srule2}, here we perform the explicit check. 

\begin{figure}[h]
\centering 
\begin{subfigure}[b]{0.32\linewidth}
\centering
\includegraphics[scale=0.20]{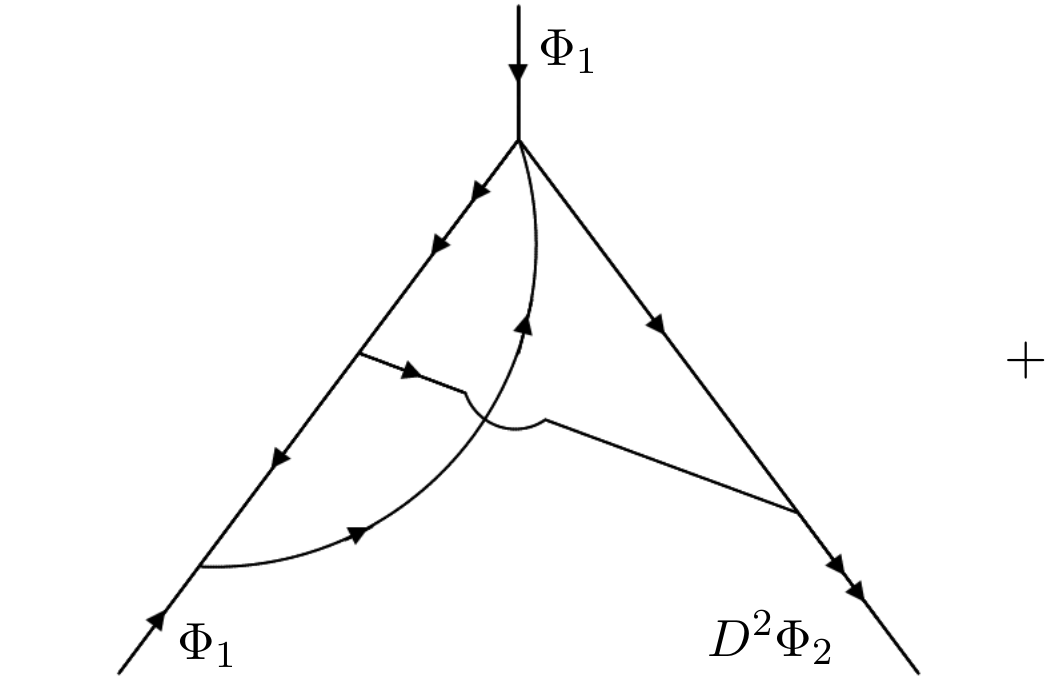}
\end{subfigure} 
\begin{subfigure}[b]{0.32\linewidth}
\centering
\includegraphics[scale=0.20]{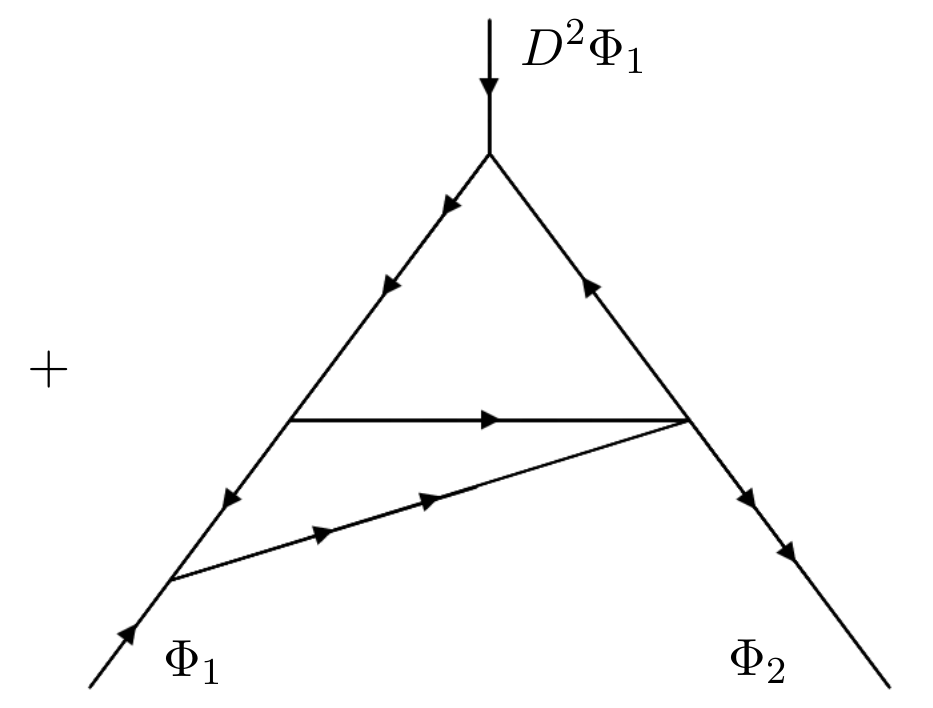}
\end{subfigure}
\centering
\begin{subfigure}[b]{0.32\linewidth}
\centering
\includegraphics[scale=0.20]{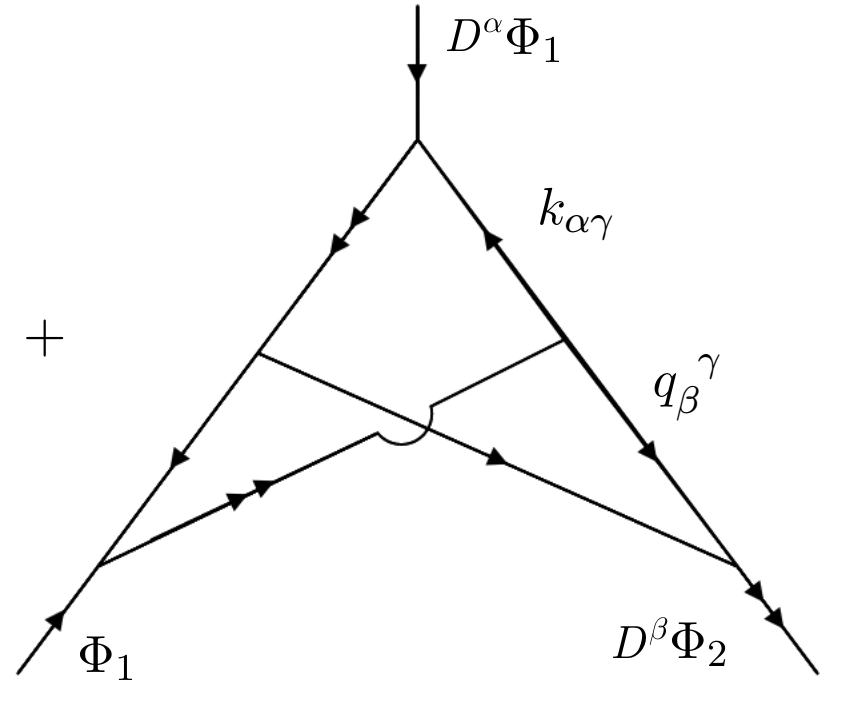}
\end{subfigure}
\caption{Diagrams resulting from the D-algebra reduction of diagram \ref{fig14_2 loop supervertici a 3}(d).}
\label{fig15_2 loop supervertici a 3_specific diagrams}
\end{figure}

For instance, focusing on the arrow configuration \ref{fig14_2 loop supervertici a 3}(d), the result of D-algebra is given in fig. 
\ref{fig15_2 loop supervertici a 3_specific diagrams}.

In the first two diagrams the covariant derivatives act on the external fields or they are responsible for the simplification of some propagators. In fact, there are some effective 4-point vertices due to Dirac $ \delta $-functions arising in this way.
In both cases we are left with a loop containing three propagators whose arrows form a closed loop, and then there is enough regularity to apply the Jordan's lemma and conclude that they vanish.

Due to the structure of the external covariant derivatives, the only relevant contribution from the third diagram in fig. \ref{fig15_2 loop supervertici a 3_specific diagrams}  is proportional to
\beq
\begin{aligned}
 & \int \frac{d \omega_q d^2q}{(2 \pi)^3} \, \frac{d \omega_k d^2k}{(2 \pi)^3} \;
\epsilon_{\alpha \beta}\left( m(\omega_k + \omega_q) + \vec{k} \cdot \vec{q} \right)
  \, \frac{1}{2m \omega_k - \vec{k}^2 + i \varepsilon} \, \frac{1}{4m (\omega_{p_1} + \omega_k) - (\vec{p}_1 + \vec{k})^2 + i \varepsilon} \\
& \qquad  \qquad \qquad  \qquad\times  \frac{1}{2m (\omega_{k} + \omega_q - \omega_{p_2}) - (\vec{k} + \vec{q} - \vec{p}_2)^2 + i \varepsilon}
 \, \frac{1}{4m (\omega_{k} + \omega_q) - (\vec{k} + \vec{q})^2 + i \varepsilon}  \\
& \qquad  \qquad \qquad  \qquad\times  \frac{1}{2m \omega_{q}  -\vec{q}^2 + i \varepsilon}
 \, \frac{1}{2m (\omega_{p_1} + \omega_{p_2}- \omega_q) - (\vec{p}_1 + \vec{p}_2 - \vec{q})^2 + i \varepsilon}  
\end{aligned}
\eeq
where momenta $ (\omega_{p_a}, \vec{p}_a) $, $a=1,2$ refer to the external $\Phi_1, \Phi_2$ particles. 
At the numerator we have used the null reduction of the 4d expression $k_{\alpha \dot{\alpha}} \, q_{\beta}^{\,\,\, \dot{\alpha}} =  (\sigma^M)_{\alpha \dot{\alpha}} (\sigma^N)_{\beta}^{\,\,\, \dot{\alpha}} k_M q_N $.
 
If we now focus on the $\omega_k$ integration, we see that in the region of large $ \omega_k $ the worst integrand goes as $1/\omega_k^3$. This allows to apply Jordan's lemma and compute the integral by residue theorem. Since all the poles are on the same side of the complex plane the result is zero. 

The same pattern occurs for the other configurations of arrows in fig. \ref{fig14_2 loop supervertici a 3}(a)-(c). This provides a check of selection rule \ref{srule2} in this particular case.



\end{document}